\documentclass[10pt,preprint]{aastex}
\newcommand{\hdf}{HDF--N}
\newcommand{\hst}{\textit{HST}}
\newcommand{\wfu}{\hbox{$U_{300}$}}

\newcommand{\wfv}{\hbox{$V_{606}$}}

\newcommand{\nicj}{\hbox{$J_{110}$}}
\newcommand{\nich}{\hbox{$H_{160}$}}

\newcommand{\AAA}{\hbox{\AA}}

\newcommand{\lsim}{\lesssim}
\newcommand{\gsim}{\gtrsim}

\newcommand{\etal}{et al.}
\newcommand{\eg}{e.g.}
\newcommand{\ie}{i.e.}

\newcommand{\kms}{\hbox{km~s$^{-1}$}}

\newcommand{\uit}{\textit{UIT}}
\newcommand{\uitastro}[1]{\hbox{ASTRO--{#1}}}

\newcommand{\kmsmpc}{\hbox{km~s$^{-1}$~Mpc$^{-1}$}}

\slugcomment{Accepted for Publication in the Astrophysical Journal}

\shorttitle{THE INTERNAL UV--TO--OPTICAL COLOR DISPERSION}
\shortauthors{PAPOVICH ET AL.}

\begin{document}

\title{THE INTERNAL ULTRAVIOLET--TO--OPTICAL COLOR DISPERSION: QUANTIFYING THE MORPHOLOGICAL $K$--CORRECTION}	

\author{\sc Casey Papovich\altaffilmark{1}, 
	Mauro Giavalisco\altaffilmark{2},
	Mark Dickinson\altaffilmark{2,3},
	Christopher J. Conselice\altaffilmark{4},
	and Henry C. Ferguson\altaffilmark{2,3}}

\altaffiltext{1}{Steward Observatory, The University of Arizona,  933
	N. Cherry Avenue, Tucson, AZ 85721; 
	\texttt{papovich@as.arizona.edu}} 
\altaffiltext{2}{Space Telescope Science Institute, 3700 San Martin
	Drive, Baltimore, MD 21218; \texttt{mauro, med,
	ferguson@stsci.edu}} 
\altaffiltext{3}{Deparment of Physics and Astronomy, The Johns Hopkins
	University, Baltimore, MD 21218}
\altaffiltext{4}{Palomar Observatory, California Institute of
	Technology, MS 105--24, Pasadena, CA 91125;
	\texttt{cc@astro.caltech.edu} }


\begin{abstract}

We present a quantitative measure of the  internal color dispersion
within galaxies, which quantifies differences in galaxy
morphology as a function of observed wavelength.  We apply this
statistic to a sample of local galaxies with archival images at
1500~\AA\ and 2500~\AA\ from the \textit{Ultraviolet Imaging
Telescope}, and ground--based $B$--band observations in order to
investigate how the internal dispersion between these colors relates
to global galaxy properties (\eg, luminosity, color, morphological
type).  In general, the dispersion in the internal galaxy colors
correlates with transformations in the galaxy morphology as a function
of wavelength, \ie, our internal color dispersion statistic quantifies
the morphological $K$--correction.  Mid--type spiral galaxies exhibit
the highest dispersion in their ultraviolet--to--optical internal
colors, which stems from differences in the stellar content that
constitute the bulge, disk, and spiral--arm components.  Irregulars
and late--type spirals show moderate internal color dispersion,
although with lower values relative to the mid--type spirals.  This
implies that young stars generally dominate the
ultraviolet--to--optical galaxy colors, modulo variations in the dust,
gas, and stellar distributions.  Ellipticals, lenticulars, and
early--type spirals generally have low or negligible internal color
dispersion, which indicates that the stars contributing to the
ultraviolet--to--optical emission have a very homogeneous distribution.
We discuss the application of the internal color dispersion to
high--redshift galaxies in deep, \textit{Hubble Space Telescope}
images.   By simulating the appearance of the local galaxy sample at
cosmological distances, many of the galaxies have luminosities that
are sufficiently bright at rest--frame optical wavelengths to be
detected within the limits of the currently deepest near--infrared
surveys even with no evolution.  Under assumptions that the luminosity
and color evolution of the local galaxies conform with the measured
values of high--redshift objects, we show that galaxies' intrinsic
internal color dispersion remains measurable out to $z\sim 3$.

\end{abstract}
 
 

\section{Introduction}

The distribution of galaxy morphology in the local Universe contains
the imprints of the processes of galaxy formation and evolution.
Indeed, the fact that galaxy morphology correlates with physical
parameters (such as optical size, luminosity, surface brightness,
total stellar/gas--mass, etc.) implies that the stages of the
morphological sequence are intertwined with the assembly of the
galaxies' stellar populations \cite[\eg][]{rob94}.  Locally, galaxies
are classifiable in terms of the well--known ``Hubble Sequence'',
which seems to apply even to moderately high redshifts ($z \sim 1$;
Schade \etal\ 1995; Lilly \etal\ 1998; Dickinson 1999; Simard \etal\
1999; van den Bergh, Cohen, \& Crabbe 2001).  At still higher
redshifts ($z\gsim 1.5$), galaxies appear to undergo a
transformation.  Galaxies at these epochs are observed to have highly
irregular, or centrally compact morphologies \citep[\eg,][]{gia96},
and the Hubble Sequence seems to no longer apply \citep{vdb02}.
However, galaxy morphological classification has been calibrated by
their appearances at rest--frame optical wavelengths, which, for the
case of the high--redshift galaxies, is only possible using
near--infrared images. To investigate the evolution of galaxy
morphology at high redshifts using optical--to--near-infrared surveys,
one must comprehend how the appearance of galaxies at rest--frame
ultraviolet (UV)--to--optical wavelengths relates to the processes of
galaxy evolution.

The degree to which the observed rest--frame UV morphology of a
galaxy corresponds to that in the rest--frame optical is a function
of galaxy spectral type.  Recent work has broadly concluded that a
galaxy appears as a later--type system when observed in the UV
relative to the optical (\ie, the morphological $K$--correction; see
Bohlin \etal\ 1991; Giavalisco \etal\ 1995; Marcum \etal\ 2001;
Kuchinski \etal\ 2000, 2001; Windhorst \etal\ 2002, and references therein).
This effect is attributed to the fact that the rest--frame far--UV
(FUV, $\sim 1500$~\AA) light stems primarily from the young O-- and B--type
stars, whereas stars of later type (A--type and later) contribute
increasingly at mid--UV (MUV, $\sim 2500$~\AA) and redder wavelengths.
Because bright, early--type stars have short relative lifespans, FUV
observations are generally sensitive to recent star formation events,
while longer--lived, later--type stars trace past star--formation
episodes, now only evidenced by an older stellar population.  Young,
bright OB--star complexes tend to dominate the spectral energy
distributions (SEDs) at UV--to--optical wavelengths in late--type
spiral and irregular galaxies that are actively forming stars.  As
such, the morphological appearances of these galaxies exhibit only a
weak (or nonexistent) dependence on UV/optical wavelength.  Conversely, the morphological $K$--correction
can be much more pronounced in early--to--mid-type spirals.  Older
stars dominate the optical portion of the SED in these galaxies, and
the UV emission originates in rather sparse regions of ongoing
star--formation, normally confined to the spiral arms.  When observed
at UV wavelengths, light from the older stars disappears and the
galaxy takes on the appearance of a later--type system.  Lastly, it is
noteworthy that elliptical and lenticular galaxies usually exhibit
weak morphological $K$--corrections.  The distribution of stellar
populations in these systems is generally very homogeneous.  These
galaxies typically have low specific star--formation rates (\ie,
star--formation rate per unit stellar mass), and any UV emission in
present--day early--type galaxies usually stems from late evolutionary
stages of late--type stars that are well mixed with the underlying
stellar populations\footnote{In a certain phase of the
post--Main--Sequence evolution, later--type Horizontal--Branch (HB)
stars can eject the bulk of their photospheres, forming planetary
nebulae, and thus expose the hot, bright cores, which do radiate in
the FUV.  These stars are indicative of  older stellar populations
(ages greater than several Gyr)  and are responsible for producing
this ``UV--upturn'' (also called the UV excess) in early--type
galaxies \citep[\eg][]{dor95,bro00}. Note that such galaxies have low
FUV--to--optical flux ratios, $m_{1500} - V \sim 5-7$ (Dorman \etal\
1995; converted to the AB magnitude system, see
\S~\ref{section:data}), which is much weaker than the values for
star--forming galaxies. \label{footnote:uvupturn}}, although certain
examples of early--type galaxies with significant star formation do
exist and appear as later--type systems when viewed at rest--frame UV
wavelengths \citep[see, \eg,][]{mar01}.

Many investigations of the galaxy morphological evolution have focused
on the relationship between structural parameters and quantitative
indexes as a function of galaxy luminosity, colors, and Hubble type
\citep[\eg,][]{sch95, abr96, abr99, van96, low97, bri98, mar98, lil98,
sim99, con00, men01}.  However, studying this evolution as a function
of redshift is hampered due to effects arising from the morphological
$K$--correction.  For example, \citet{gia95} artificially redshifted
local galaxies, spanning a range of Hubble type using FUV images from
the \textit{Ultraviolet Imaging Telescope} (\uit), and demonstrated
that these broadly reproduce the observed morphology of the faint
galaxies in optical images from the \textit{Hubble Space Telescope}
\citep[\hst; see also][]{boh91}.  Thus, Giavalisco \etal\ suggested
that dynamical evolution may only have played a minor role in the
galaxies' evolutionary histories and argued caution when interpreting
the faint--blue--galaxy population in the context of galaxy--evolution
scenarios.  These conclusions have been elaborated upon by
\citet{kuc01}, who preformed very detailed simulations of artificially
redshifting \uit/FUV images of a larger sample of local galaxies into
the \hst/WFPC2 data for the Hubble Deep Field North
\citep[HDF--N;][]{wil96}.  Kuchinski \etal\ concluded that effects
arising from bandshifting dominate the morphological $K$--correction,
and surface--brightness dimming induces small (but non-negligible),
seemingly random changes into quantitative morphological parameters.
Thus, comparing the rest--frame UV morphologies of the high--redshift
galaxy population to those in optical images of local galaxies does
not likely provide a homologous comparison sample.

Additional insights --- beyond those possible from single--band
morphology studies --- can be obtained by investigating the
distribution and morphology of internal colors and color gradients
within individual galaxies.  In a novel study, \citet{dej96} showed
that the color gradients observed in local spiral galaxies are likely
the result of a combination of stellar age and metallicity effects,
with the outer disk regions being bluer and having lower metallicity.
\citet{abr99} and Menanteau \etal\ (2001) investigated the internal colors of
galaxies in the HDF--N and HDF--S using the \hst\ WFPC2 four--color
data to study variations in the star--formation histories of field
galaxies.   For example, Menanteau \etal\ observed a significant
fraction of galaxies in the field population with strong internal
color variations, which they interpreted as evidence for recent star
formation and argued that (field) elliptical and bulge--dominated
galaxies have continued to assemble their stellar populations since
$z\sim 1$.  In contrast, elliptical galaxies in rich clusters exhibit
relatively lower internal color variation, which argues that color
variations in early--type galaxies have a strong environmental
dependence, where evolution is accelerated in the densest environments
(\ie, enhanced star formation at earlier epochs).   The broad
conclusion from these studies is that the synthesis of galaxy internal
colors provides insights into the spatial distribution of the
underlying stellar populations, and thus into the processes of star
formation in the galaxies \citep[see
also,][]{bel01b,gad01,esk03}.

In this paper, we develop a statistic to measure the morphological
difference between galaxy images in two bandpasses.  As reviewed
above, the wavelength dependence of morphology has often been
previously described in qualitative terms.  Here, we aim to provide a
simple, \textit{quantitative} measure of morphological difference
between images at different wavelengths that can be applied, for
example, to \hst\ images of distant galaxies.  Our statistic
quantifies the amplitude of the morphological differences or the
variations in internal galaxy colors, without describing the spatial
organization of those variations.  Some other measurements, such as
color gradients (see references above), emphasize the organization of
color differences.  Ours is, in a sense, cruder than a structural
quantity like a color gradient, but it can be measured in a simple,
automated, and model independent fashion, and provides a
straightforward measure of the degree to which stellar--population or
extinction variations within a galaxy affect the relative morphologies
seen at two wavelengths.  Furthermore, because our statistic is
integrated over a whole galaxy, it is robustly measurable for even
faint, distant galaxies with relatively low, signal-to-noise ratios.

Here, we focus on the internal color dispersion for a sample of local
galaxies using \uit/UV and ground--based optical data, and compare
this to their morphological features and integrated properties.    In
a forthcoming paper (Papovich \etal, 2003; Paper II), we apply this
diagnostic to samples of high--redshift galaxies at $z\sim 1$ and
$z\sim 2.7$ using the \hst\ UV--to--near-infrared data for the \hdf.
An advantage of using the \uit\ and $B$--band data for the local
galaxy sample is that they probe nearly equal rest--frame wavelengths
as the \hst\ data available for the high--redshift \hdf\ galaxies.
Therefore, we can use the results here as a baseline for
interpretation of the processes affecting the observations of the
higher redshift galaxies.

It is useful at this point to postulate various properties that could
produce or suppress a galaxy's internal color dispersion.  Because
we  focus on UV/optical wavelengths, we expect that the internal
color dispersion should be sensitive to the degree of heterogeneity in
the relative composition and spatial distribution of a galaxy's young
and more aged stars.   Small values of the internal color dispersion
could indicate several scenarios.  For example, all stars in a galaxy
could be old (with little UV emission), with small scatter in their
ages and colors.  Alternatively, either a coeval stellar population
could dominate the light for all UV/optical wavelengths, or the galaxy
could contain a mixed--aged stellar population with a homogeneous
distribution, \eg, the stars could be well--mixed following a merger
event, or after several dynamical timescales within a galactic disk.
These scenarios likely apply to ellipticals, lenticulars, and
early--type spirals, and thus one may expect these types of objects to
exhibit low internal color dispersion.  Small dispersion in a galaxy's
internal colors could also signify that a substantial young stellar
population with fairly uniform colors dominates the UV/optical galaxy
colors, with the light from other (older) stellar populations lost in
the ``glare'' of the nascent stars.  This case is arguably relevant
for late--type irregular and starbursting galaxies.  Conversely, a
large internal color dispersion in these colors probably requires both
a range of stellar population ages (or a highly inhomogeneous
distribution of dust extinction) and spatial inhomogeneity.  This
description plausibly predicts that mid--type spirals --- with
non--coeval stellar populations in the bulge/disk/spiral--arm
components --- should display the greatest internal UV/optical--color
dispersion.  Therefore, our definition of a galaxy's internal color
dispersion has the potential to quantify the prominence of these
morphological features, and should vary as a function of morphological
type.

The outline for the remainder of this article is as follows. In \S2,
we discuss the data used and the selection of the sample of local
galaxies.  In \S3, we introduce the internal color dispersion
statistic and demonstrate its robustness against changes in object
signal--to--noise ratio (S/N).  We then discuss the application of
this measure to the local galaxy sample.  In \S4, we compare the
measured internal color dispersion values to other galaxy properties
to investigate those processes to which this metric is sensitive.  In
\S5, we discuss the application of this diagnostic to high--redshift
galaxy populations by artificially redshifting galaxies from the local
sample and inserting them into deep \hst\ images.  Finally, in \S6, we
present a summary of this work.  Unless otherwise specified, all
magnitudes in this paper are presented on the AB system \citep{oke96},
where $m(\mathrm{AB}) = 31.4 - 2.5 \log(f_\nu/\mathrm{nJy})$.
Throughout this work we use a flat, cosmological--constant--dominated
world model, $\Omega_m = 0.3$, $\Lambda=0.7$, and a Hubble constant of
$H_0 = 70$~\kmsmpc.


\section{The Data and Galaxy Sample\label{section:data}}

We have selected nearby galaxies for which there exist archival
imaging available at FUV, MUV, and optical (\ie, $B$--band)
wavelengths.  The FUV and MUV data used here were collected by the
\uit\ aboard the \uitastro{1}\ space--shuttle mission in
1990.\footnote{All \uit\ data were obtained from the Multimission
Archive at Space Telescope (MAST), \texttt{http://archive.stsci.edu}}
The \uit\ mission included two cameras sensitive to a combined
wavelength range roughly $1200-3300$~\AA\ (see Stecher \etal\ 1997 for
a presentation of the characteristics of the \uit\ instrument and data
reduction/calibration).  We have restricted our analysis to those UV
images obtained from the two broad \uit\ bandpasses: FUV, \uit/B1,
$\lambda_\mathrm{eff} = 1521$~\AA, $\Delta\lambda = 354$~\AA; MUV,
\uit/A1, $\lambda_\mathrm{eff} = 2488$~\AA, $\Delta\lambda =
1147$~\AA.  The \uit\ instrument was flown on a second shuttle
mission in 1995, \uitastro{2}, but the MUV camera failed to activate
on orbit.  Thus, only FUV data is available from the later mission.
To augment the available FUV and MUV data from the \uitastro{1}
mission, we have made use of publicly available ground--based optical
images obtained in support of the \uit\ mission by \citet{che96} from
telescopes at the Kitt Peak National Observatory, Cerro Tololo
Inter--American Observatory, and Mount Laguna Observatory.  The final
\uit\ dataset includes 14 local galaxies that represent the range
of morphological types of the local galaxy population (see
table~\ref{table:localprop}).

All the \uit\ images have been linearized and flat fielded as
described in \citet{ste97}.  In most cases, the data have also been
corrected for geometric distortions caused in the image phototubes.
For a few of the objects in our sample, a distortion--corrected image
was not available from the \uit\ archive.  For these images, we have
applied a distortion correction using the IDL routine DISTWARP, which
is included in the \uit--analysis software, MOUSSE, available on the
world--wide
web.\footnote{\texttt{ftp://archive.stsci.edu/pub/astro/uit/software/}}
After data processing and distortion correction, the final images have a
pixel scale of $1\farcs12$~pixel \citep[see][]{ste97}.  The \uit\
images are marginally oversampled:  the image point--spread functions
(PSFs) have a full--width at half maximum (FWHM)  of
$\mathrm{FWHM(FUV)} = 3\farcs0$ and $\mathrm{FWHM(MUV)} = 2\farcs7$
for stars within $< 16$\arcmin\ of the image center \citep{ste97}.
Many of the \uit\ images suffer from scratches and other cosmetic
defects from the photographic development.  These were removed by hand
using the IRAF\footnote{The Image Reduction and Analysis Facility
(IRAF) software is provided by the National Optical Astronomy
Observatories (NOAO), which is operated by the Association of
Universities of Research in Astronomy for Research in Astronomy, Inc.,
under contract to the National Science Foundation.} routine IMEDIT.
The archival \uit\ images have been calibrated in units of flux per
unit wavelength, \ie, erg~s$^{-1}$~cm$^{-2}$~\AA$^{-1}$, which
correspond to ``ST'' magnitudes (\ie, magnitudes based relative to a
``flat'' reference spectrum in flux per unit {\AA}ngstrom),
$m(\mathrm{ST}) = -2.5\, \log(f_\lambda) - 21.1$.  We have converted
these to AB magnitudes (\ie, magnitudes based relative to a ``flat''
reference spectrum in flux per unit frequency) by comparing the
magnitudes derived for a flat spectrum ($f_\nu \sim \nu^0$) source
through the \uit\ bandpasses, using the IRAF/STSDAS\footnote{The Space
Telescope Science Data Analysis System (STSDAS) is distributed by the
Space Telescope Science Institute.} SYNPHOT package.  The resulting
conversions are $m(\mathrm{AB}) \simeq m(\mathrm{ST}) + 1.7$~mag for
\uit/A1 photometry, and $m(\mathrm{AB}) \simeq m(\mathrm{ST}) +
2.8$~mag for \uit/B1 photometry.  Our tests have shown that adopting
other plausible reference spectra (either with power--law exponent
$f_\nu \sim \nu^{\alpha}$ for $-2 \leq \alpha \leq 2$ or using the
spectrum of a Vega--like star) causes negligible changes on these
conversion factors (\ie, $< 1$\%).

The ground--based $B$--band images were processed by \citet{che96}.
Images were then flux--calibrated using standard stars taken on or
near the nights of the observations.  Magnitudes were converted to the
AB magnitude system by comparing a flat spectrum source to that of
Vega through a $B$--band passband again using the SYNPHOT
package.  In both the ground--based and \uit\ images, many of the
objects contain obvious foreground stars, which were masked out
and replaced by background sky pixels using the IRAF task
IMEDIT.

The \uit\ images were registered to the ground--based images
using point sources common between the UV and optical images.  In
several instances, the images are devoid of common point sources, in
which case we aligned images using the astrometric solutions included
in the image headers.  All registrations were subsequently inspected
by eye to ensure that similar features of the galaxies in the sample
are matched between the final images.  Slight misalignments (ranging
from a fraction of a pixel to a few pixels) were detected and
corrected while comparing the galaxy internal colors.

To investigate the quantitative morphology of galaxies with data taken
through various bandpasses, it is essential that the images be matched
to a common PSF.  This will reduce or eliminate image--to--image
variations between bandpasses resulting purely from PSF differences.
For the \uit\ and ground--based data, we fit Moffat--profile PSFs to
stars identified in the images.  Moffat profiles include a term to
increase the strength of the kernel in the ``wings'' of the PSF
relative to a pure Gaussian.  This is appropriate for ground--based
observations as these profiles match the seeing caused by atmospheric
turbulence and other observational effects \citep{ben90}.  The Moffat
profile also provides an acceptable fit to the \uit\ PSF, which
contains broader wings over a Gaussian profile (see Stecher \etal\
1997).   Many of the \uit\ B1 (FUV) images lacked point sources
(as only relatively rare O and B stars are bright for $\lambda \lsim
1700$~\AA).  For these images, we used a moffat--profile fit to the
average UIT/B1 PSF described in \citet{ste97}.  For each object, we
convolved the ground--based images with the \uit\ PSFs and vice versa
in order to match the PSFs between the various images.  Inspecting
stars in each of the PSF--matched images, the flux profile contained
by individual point sources agrees very well, with $\lsim 10$\%
difference out to radii $\approx 10-20$\arcsec.

In table~\ref{table:localprop}, we present flux and structural
properties for the 14 galaxies in the \uit\ sample.  Columns (1) and
(2) indicate the primary name (\eg, NGC~ID number) and any alternate
name, respectively.  The galaxy morphological type, and velocity in
the Galactic standard of rest ($V_\mathrm{GSR})$, taken from the
\textit{Third Reference Catalogue of Bright Galaxies}
\citep[RC3;][]{dev91}, are given in columns (3) and (4).  Columns (5)
and (6) indicate the assumed distance to each galaxy and associated
reference.  In columns ($7-10$), we give the apparent $B$--band
magnitude, absolute $B$--magnitude [using the distance measure from
column (5)], and $m_{2500} - B$, $m_{1500} - B$ colors,
measured from elliptical apertures constructed from the galaxies'
$B$--band radial profiles \citep{kro80} using the SExtractor software
\citep[v2.2.2; the MAG\_AUTO photometry, see][]{ber96}.  All
the magnitudes given in table~\ref{table:localprop} for each bandpass
have been corrected for Galactic extinction using the color--excess
maps of \citet{sch98} and the Galactic extinction curve of
\citet{car89}; we list the corresponding extinction in the $B$--band
in column (11).  The galaxy photometry and colors derived here are
generally consistent   with others reported in the literature, modulo
differences in the Galactic extinction used \citep[for example,
cf.,][]{mar01}.  

\begin{deluxetable}{clcccccccccc}
\rotate
\tabletypesize{\small}
\tablewidth{0pt}
\tablecaption{Properties of the Local Galaxy Sample
\label{table:localprop}}
\tablehead{ \colhead{} & \colhead{} & \colhead{} &
\colhead{} & \colhead{$V_\mathrm{GSR}$\tablenotemark{1}} &
\colhead{Dist.} & \colhead{Dist.} & \colhead{$m(B)$} &
\colhead{$M(B)$} & \colhead{} & \colhead{}
& \colhead{$A_\mathrm{g}(B)$} \\
\colhead{} & \colhead{Name} & \colhead{Alt.\ Name} &
\colhead{Morph.\ Type\tablenotemark{1}} & \colhead{[\kms]} &
\colhead{[Mpc]} & \colhead{Ref.} & \colhead{[mag]} &
\colhead{[mag]} & \colhead{$m_{2500}\!\!-\!\!B$} & \colhead{$m_{1500}\!\!-\!\!B$}
& \colhead{[mag]} \\
\colhead{} &  \colhead{(1)} & \colhead{(2)} &
\colhead{(3)} & \colhead{(4)} &
\colhead{(5)} & \colhead{(6)} & \colhead{(7)} &
\colhead{(8)} & \colhead{(9)} & \colhead{(10)}
& \colhead{(11)}}
\startdata
\phn1 & NGC~628 &  M74 & SA(s)c       & \phn753 & \phn7.3 & 1 & \phn9.23 & -20.1
& 2.3 & 3.0 & 0.301 \\
\phn2 & NGC~1068 & M77 & SA(rs)b & 1144    & 16.3    & 2 & \phn9.61 & -21.5
& 2.4 & 3.0 & 0.145 \\
\phn3 & NGC~1275 & Perseus A & E0, pec & 5362 & 76.6 & 2 & 12.09 &
-22.3 & 2.1 & 2.2 & 0.703 \\ 
\phn4 & NGC~1316 & Fornax A & SAB(s)0, pec & 1674 & 15.9 & 3 & \phn9.82 &
-21.2 & 3.9 & 6.6 & 0.089 \\ 
\phn5 & NGC~1317 & Fornax B & SAB(r)a      & 1822 & 15.9 & 3 & 12.01 & -19.0 &
3.2 & 3.5 & 0.089 \\
\phn6 & NGC~1399 & \nodata& E1, pec & 1323 & 15.9	& 3 & 10.51 & -20.5 & 3.8 &
5.0 & 0.056 \\
\phn7 & NGC~2146 & \nodata& SB(s)ab, pec & 1035 &	14.8 & 2 & 11.17 & -19.7 &
2.8 & 5.0 & 0.415 \\
\phn8 & NGC~2992 &\nodata& Sa, pec & 2125 & 31.1 & 2 & 12.67 & -19.8 &
2.9 & 2.6 & 0.261 \\
\phn9 & NGC~2993 & \nodata& Sa, pec      & 2224 & 31.1 & 2 & 12.81 & -19.7 &
0.8 & 1.5 & 0.261 \\
10 & NGC~3031 & M81  & SA(s)ab & \phn\phn69 &	\phn3.5 & 4 & \phn7.67 & -20.0
& 3.0 & 4.4 & 0.346 \\
11 & NGC~3034 & M82  & I0        & \phn323 & \phn3.5 & 4 & \phn8.90 & -18.9
& 2.7 & 5.1 & 0.685 \\
12 & UGC~06697 & \nodata & Im           & 6697 & 95.7 & 2 & 14.00 & -20.9 &
0.9 & 1.3 & 0.094 \\
13 & NGC~4321 & M100 & SAB(s)bc     & 1540 & 14.1 & 5 & 10.30 & -20.4 &
2.4 & 2.8 & 0.113 \\
14 & NGC~4486 & M87  & E0, pec & 1229 & 21.1 & 6 & \phn9.53 & -22.1 &
5.0 & 7.9 & 0.096 \\
\enddata
\tablenotetext{1}{de~Vaucouleurs \etal\ (1991), \textit{Third
Reference Catalogue of Bright Galaxies} (RC3).}
\tablerefs{1.\ Sharina, Karachentsev, \& Tikhonov 1996.  2.\ Distance
  computed from the velocity relative to the Galactic standard rest
  frame (column 4) and assuming a Hubble constant of 70~\kmsmpc.  3.\
  Fornax cluster members: Freedman \etal\ 2001 (see also Kohle \etal\
  1996; Shaya \etal\ 1996; McMillan, Ciardullo, \& Jacoby 1993).  4.\
  Freedman \etal\  1994, 2001. 5.\ Ferrarese \etal\ 1996, Freedman
  \etal\  2001.  6.\ Gavazzi \etal\ 1998.  }
\end{deluxetable}


\section{Galaxy Internal Colors}

\subsection{Internal Color Dispersion\label{section:icd}}

To investigate the two--dimensional color structure of the galaxies,
we have developed a  flux--independent statistic capable of
incorporating information on the internal color dispersion of a galaxy
about its mean color.   Here, we define the internal color dispersion
as the ratio of the  squared difference of image flux--intensity
values about the mean galaxy color to the square of the total image
flux, \ie,
\begin{equation}\label{eqn:rawicv}
\xi(I_1,I_2) = \frac{ \sum ( I_2 - \alpha\, I_1 - \beta)^2}
	{\sum (I_2-\beta)^2},
\end{equation}
where $I_1$ and $I_2$ are the image flux--intensity values for each
object obtained in each of two passbands.   The scaling factor
$\alpha$ is the flux ratio, or {\it integrated color}, between images
$I_2$ and $I_1$, while the linear offset $\beta$ adjusts (if
necessary) for differences in the relative background levels of the
two images.  In practice, we calculate these terms by minimizing the
statistic, $\chi^2 = \sum [ (I_2 - \alpha\, I_1 - \beta) / \sigma
]^2$, adjusting $\alpha$ and $\beta$ as free parameters (where
$\sigma$ represents the uncertainties on the image flux--intensity
values).

To compute $\xi$, we sum over all pixels within in single radius.
Galaxy radii defined using limiting surface--brightness criteria
(e.g., the Holmberg radius or the limiting isophote where the galaxy
surface--brightness equals the sky brightness) are poorly defined when
trying to compare a wider range of galaxy types especially at various
redshifts.  Here, we define the aperture size as $1.5\times r_p(\eta =
0.2)$, where $r_p$ is the  Petrosian radius of $I_1$ as defined for
the \textit{Sloan Digital Sky Survey} \citep[see][and references
therein]{bla01}, and $\eta(r)$ is defined as the ratio of the galaxy
surface brightness, $I(r)$, averaged over an annulus of radius $r$, to
the mean surface brightness within this radius, $\langle I(r)
\rangle$.  The Petrosian radius only depends on the
surface--brightness distribution of the galaxy and is thus independent
of redshift or systematics in the image calibration.  The chosen
aperture encompasses most of the light from the galaxy without adding
excessive background light, and has been used to compute other
quantitative morphological parameters \citep[see, \eg,][]{con00,con03}.

The definition in equation~\ref{eqn:rawicv} provides an incomplete
description as it neglects the contribution to the dispersion arising
solely from the background pixel intensity distribution (\ie, the
image sky ``noise''). For example, given two images with mean pixel
value zero, but rms uncertainties, $\sigma_1$ and $\sigma_2$,
respectively, and an arbitrary scaling factor $\alpha$,
equation~\ref{eqn:rawicv} has a net, non--zero value, $(\sigma_2^2 +
\alpha^2\, \sigma_1^2)/\sigma_2^2$.  In theory the statistical sky
contribution can be removed if one knows the exact rms
pixel--to--pixel variations.
However, in practice it can be difficult to compute the local sky rms
variation in the region of the galaxy in real data.  To estimate the
contribution to $\xi$ from the background, we have adopted a procedure
to compute the pixel intensity values for a blank region of sky for
each image, which is defined as an image region equal to the aperture
size used for the object with no pixel values above the detection
threshold (\ie, an image section where all pixels have a zero value in
the SExtractor SEGMENTATION image).  One could use randomly--selected
background pixels, which would ensure that faint, undetected objects
do not contribute to the internal color dispersion. However,  using a
region of contiguous pixels provides a better estimate of noise from
any correlation between neighboring pixels (for example, from rebinning
and/or image convolution, both of which have been done to the \uit\
sample here).  In practice, we computed values the background rms
of each image by taking the mean of the values from many ($\sim 100$)
different regions.  The appropriate values are then subtracted from
the numerator and denominator of equation~\ref{eqn:rawicv} to remove
the contribution to the color dispersion from the background,
\begin{equation}\label{eqn:zeta}
\xi(I_1,I_2) \equiv 
	\frac{ \sum ( I_2 - \alpha\, I_1 - \beta)^2 -  \sum(B_2 -
		\alpha\, B_1)^2 } {\sum (I_2-\beta)^2  - \sum
		(B_2-\alpha\, B_1)^2},
\end{equation}
where $B_1$ and $B_2$ are the pixel intensity values from the  blank
region in each image.  One can also calculate the statistical error
associated with this quantity, with the approximation that the
background pixels have a Gaussian distribution,
\begin{equation}\label{eqn:zetaerr}
\delta(\xi) = \frac{ \sqrt{2/N_\mathrm{pix}}\, \sum ( B_2^2 + \alpha^2\,B_1^2)}
	{\sum(I_2-\beta)^2 - \sum (B_2-\alpha\, B_1)^2}.
\end{equation}

\begin{figure}
\plottwo{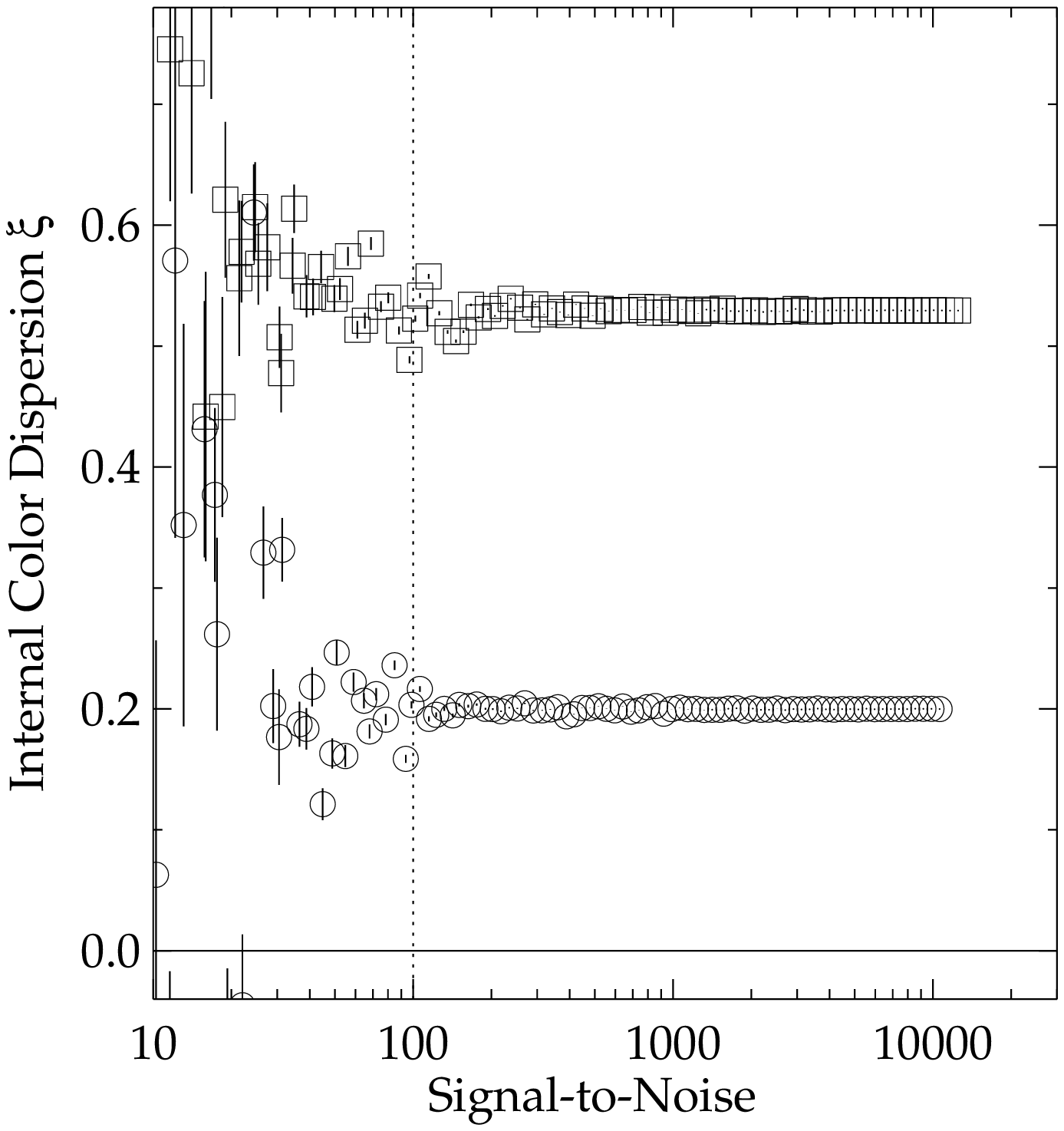}{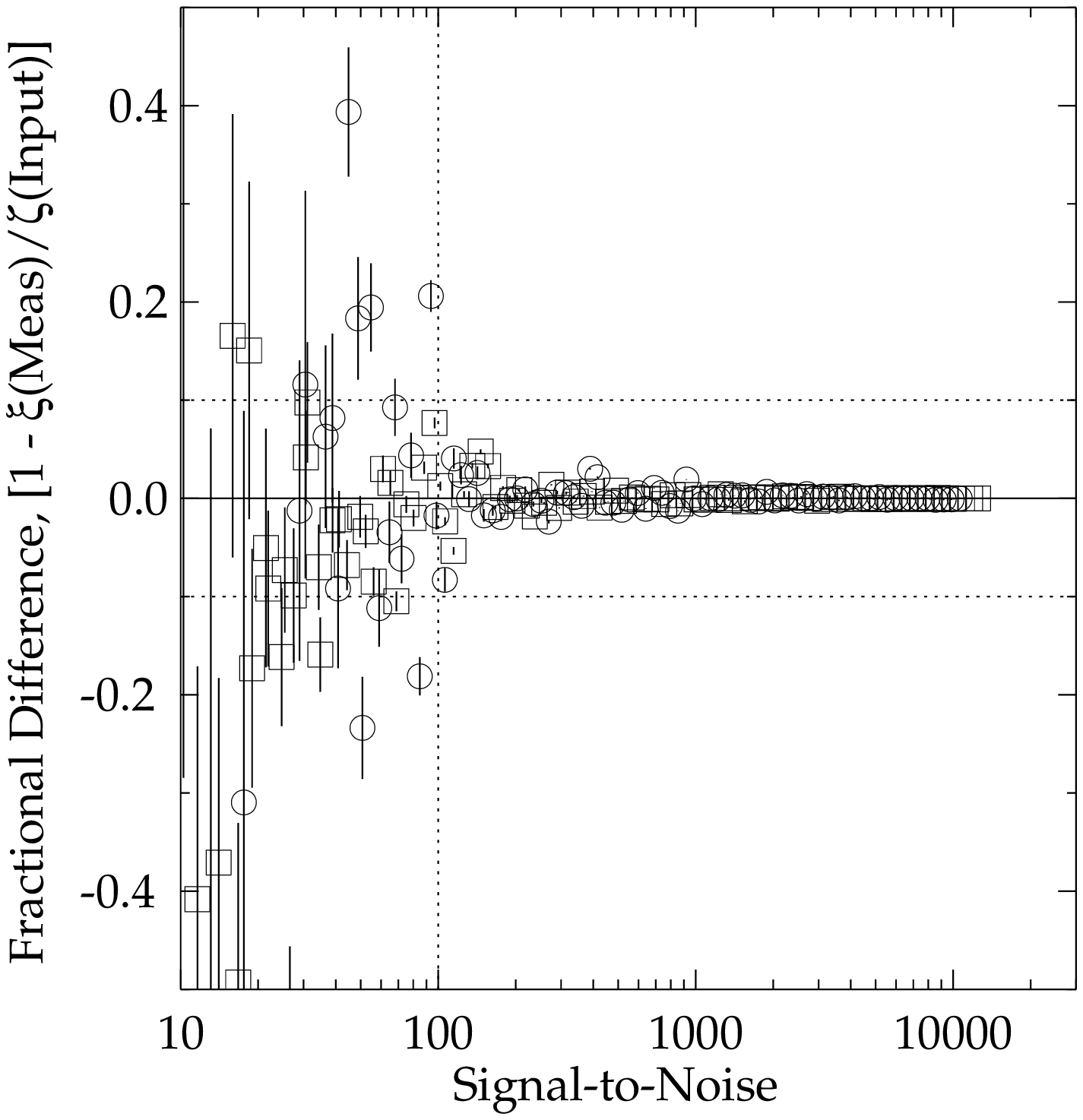}
\caption{Behavior of the internal color dispersion, $\xi$, for a two
simulated objects as a function of signal--to--noise ratio.
\textit{Left}: The panel panel shows the measured values for the
simulated objects (\textit{squares} and \textit{circles},
respectively).  \textit{Right}: The panel shows the fractional
difference, defined as $(1 - \xi(\mathrm{Measured}) /
\xi(\mathrm{Input}) )$.  For reference, the horizontal, dashed lines
illustrate a fractional difference level of 10\%.  In both panels the
vertical dotted line shows, $\mathrm{S/N} = 100$.  The value of $\xi$
seems largely unaffected by decreasing to $\mathrm{S/N} \sim 100$, and
changes by $\lsim 10$\% to $\mathrm{S/N \geq 80}$.  Below this level,
the underlying sky pixel--to--pixel rms levels are comparable to any
intrinsic signal from the galaxy, as reflected by the increasing
uncertainty in $\xi$ at low S/N.\label{fig:fakemessy}}
\end{figure}

We have tested the behavior of $\xi$ for two simulated objects with
\textit{a priori} known (fixed) total flux and internal color
dispersion.  Noise was added in increasing amounts to emulate the
effects of the sky background, and we derived $\xi$ for each simulated
object using the definition in equation~\ref{eqn:zeta}.
Figure~\ref{fig:fakemessy} shows the behavior of $\xi$ as a function
of $\mathrm{S/N} = \sum f / \sqrt{N_\mathrm{pix} \sigma^2}$, where $f$
represents the flux--intensity values, and $\sigma^2$ is the image
variance due to the noise of the background.  The sum is taken over
all pixels ($N_\mathrm{pix}$) within the object boundary.  As
illustrated in the figure, our internal--color--dispersion definition
is a robust (\ie, flux independent) statistic for objects with
$\mathrm{S/N} \gsim 100$, where uncertainties are essentially nil (see
the right panel of figure~\ref{fig:fakemessy}).  We also found that
the derived $\delta(\xi)$ values are equal to the standard deviation
of the measured $\xi$ values, which implies that the calculated
uncertainty (\ie, equation~\ref{eqn:zetaerr}) is accurate for the case
that the pixel--to--pixel noise has a Gaussian distribution (see also
the discussion in \S~\ref{sec:icd_uit}).  The internal color
dispersion remains a statistically viable, flux--independent measure
of the ``true'' value down to $\mathrm{S/N} \sim 80$, where the right
panel of fig~\ref{fig:fakemessy} shows that the fractional difference
[defined as $(\xi(\mathrm{measured}) - \xi(\mathrm{input})) /
\xi(\mathrm{input})$] of $\xi$ is $\lsim 10$\% relative to the true
value.

Our definition of $\xi$ is theoretically invariant under interchanges of
$I_1$ and $I_2$, \ie, $\xi(I_1,I_2) = \xi(I_2,I_1)$.  This is a result
of the fact that Equation~\ref{eqn:zeta} is roughly equal to the rms
of $I_2$ about a mean value (defined by $\alpha I_1$) normalized by
the squared sum of intensity of $I_2$.  Substituting $I_1$ for $I_2$
and vice versa rescales both the numerator and denominator of
Equation~\ref{eqn:zeta} by the same amount, leaving $\xi$ unchanged.
However, we have found that in practice $\xi$ is more
robust when the image with higher S/N is used as $I_1$.  Using
the lower--S/N image for $I_1$ increases the uncertainties on $\alpha$
and $\beta$ from the fitting algorithm, which translates to larger
scatter on the inferred dispersion in the internal colors.  Therefore,
we will always use the higher--S/N image as $I_1$ in
equation~\ref{eqn:zeta}.  For our analysis here, this amounts to using
the $B$--band as $I_1$  to compute $\xi(\mathrm{MUV},B)$ and the
\uit/MUV as $I_1$ to compute $\xi(\mathrm{FUV},\mathrm{MUV})$.


\subsection{Application to the \uit\ Galaxy Sample}\label{sec:icd_uit}

For each galaxy in the \uit\ sample, we derived the internal color
dispersion using equation~\ref{eqn:zeta} for the \uit/FUV, \uit/MUV,
and $B$-band images.  We have computed the internal color dispersion
between the FUV/MUV and MUV/$B$ bandpass combinations, and these
results are presented in table~\ref{table:localicd}.  We have not
included the internal color dispersion values between the \uit/FUV--
and $B$--bands as any variation in these colors includes both the
internal color variations between the FUV/MUV and MUV/$B$--bands.
Also listed in the table are the object S/N for each bandpass
(ground--based and \uit) within the isophotal apertures defined from
the $B$--band images using the SExtractor software.  To ensure that
our internal color dispersion values measure light solely from the
galaxies, we have masked out interloper objects (\eg, obvious
foreground stars) using the results in the SExtractor SEGMENTATION
image.  We then recomputed the internal color dispersion for these
galaxies using an area for the background estimate with the interloper
isophotal areas subtracted.  Four of the galaxies in the sample are
detected in the \uit/FUV with low signal--to--noise, ($\mathrm{S/N}
\leq 80$).  Thus, we will exclude these objects from the analysis of
the $\xi(\mathrm{FUV},\mathrm{MUV})$ values  (these objects are
flagged with dagger symbols in table~\ref{table:localicd}).

\begin{deluxetable}{clccccc}
\tablewidth{0pt}
\tablecaption{Derived Internal Color Dispersion of the Local Galaxy Sample
\label{table:localicd}}
\tablehead{ \colhead{} & \colhead{Name} & \colhead{$\xi(\mathrm{MUV},B)$} &
\colhead{$\xi(\mathrm{FUV,MUV})$} &
\colhead{S/N($B$)\tablenotemark{\dag}} & 
\colhead{S/N(MUV)\tablenotemark{\dag}} & 
\colhead{S/N(FUV)\tablenotemark{\dag}} \\
\colhead{} & \colhead{(1)} & \colhead{(2)} &
\colhead{(3)} & \colhead{(4)} & \colhead{(5)} &
\colhead{(6)}}
\startdata
\phn1 & NGC~628   & \phs $0.29\pm 0.00$ & \phs $0.04\pm 0.02$ &
$2.5\times10^3$ & $1.3\times 10^3$ & $4.8\times 10^2$ \\
\phn2 & NGC~1068  & \phs $0.08\pm 0.00$ & \phs $0.07\pm 0.00$ &
$3.9\times10^3$ & $1.1\times 10^4$ & $6.8\times 10^3$ \\ 
\phn3 & NGC~1275  & $-0.02\pm 0.01$ & \phs $0.07\pm 0.02$ &
$3.9\times10^3$ & $1.1\times 10^3$ & $4.2\times 10^2$ \\ 
\phn4 & NGC~1316  & \phs $0.02\pm 0.00$ & \phs $0.00\pm
0.91$\tablenotemark{\ddag}  &
$3.7\times10^3$ & $1.7\times 10^3$ & $7.6\times 10^1$ \\ 
\phn5 & NGC~1317  & \phs $0.22\pm 0.00$ & \phs $0.01\pm 0.00$ &
$3.4\times10^3$ & $7.6\times 10^2$ & $2.8\times 10^2$ \\ 
\phn6 & NGC~1399  &     $-0.01\pm 0.00$ & $-0.01\pm 0.03$ &
$1.1\times10^4$ & $1.0\times 10^3$ & $3.2\times 10^2$ \\ 
\phn7 & NGC~2146  & \phs $0.16\pm 0.01$ & \phs $0.72\pm
1.25$\tablenotemark{\ddag} & 
$1.1\times10^4$ & $4.7\times 10^2$ & $4.8\times 10^1$ \\
\phn8 & NGC~2992  & $-0.09\pm 0.18$ & $-0.17\pm
1.90$\tablenotemark{\ddag} & 
$6.3\times 10^3$ & $1.3\times 10^2$ & $5.2\times 10^1$ \\
\phn9 & NGC~2993  & \phs $0.01\pm 0.01$ & \phs $0.07\pm 0.01$ &
$6.1\times 10^3$ & $8.2\times 10^2$ & $3.0\times 10^2$ \\
   10 & NGC~3031  & \phs $0.37\pm 0.00$ & \phs $0.36\pm 0.02$ &
$3.3\times10^4$ & $3.3\times 10^3$ & $1.0\times 10^3$ \\
   11 & NGC~3034  & \phs $0.11\pm 0.00$ & \phs $1.11\pm
1.30$\tablenotemark{\ddag} & 
$8.6\times 10^3$ & $1.4\times 10^3$ & $6.4\times 10^1$ \\
   12 & UGC~06697 & \phs $0.06\pm 0.02$ & \phs $0.03\pm 0.06$ &
$4.3\times 10^3$ & $3.3\times 10^2$ & $2.7\times 10^2$ \\
   13 & NGC~4321  & \phs $0.24\pm 0.01$ & \phs $0.02\pm 0.01$ &
$4.4\times 10^3$ & $3.5\times 10^2$ & $3.4\times 10^2$ \\
   14 & NGC~4486  & $-0.01\pm 0.02$ & \phs$0.03\pm 0.01$ &
$5.2\times 10^3$ & $2.0\times 10^2$ & $1.3\times 10^2$ \\
\enddata
\tablenotetext{\dag}{Signal--to--noise ratio measured within an
isophotal aperture defined from the $B$--band image.}
\tablenotetext{\ddag}{Object is detected in \uit/1500~\AA\ with
$\mathrm{S/N}(\mathrm{FUV}) \leq 80$.}
\end{deluxetable}

The analytic uncertainty on $\xi$ (equation~\ref{eqn:zetaerr}) was
derived assuming a Gaussian distribution of background
pixel--to--pixel noise.  However, for the data used here this may be
an oversimplification.  The \uit\ data have low image backgrounds for
which a Gaussian distribution may be inappropriate.  Furthermore,
because we have resampled and convolved the data to match the pixel
scales and PSF between the \uit\ and $B$--band images, the noise
between neighboring pixels is correlated; the pixel--to--pixel random
noise is suppressed by a factor equal to $(\sum C_{ij}^2)^{-1/2}$,
where $C_{ij}$ are the elements of the correlation matrix.   Given
these systematic uncertainties in the images backgrounds, we have
resorted to Monte Carlo simulations to quantify the uncertainties in
the internal color dispersion between the \uit\ and $B$--band data.
For each galaxy in the \uit\ sample, we repeatedly inserted simulated
objects with \textit{a priori} known internal color dispersion into
blank image regions and measured the resulting scatter in the derived
$\xi$ values.   We then take the standard deviation of the measured
$\xi$ values of the Monte Carlo simulations as the uncertainty for $\xi$.
These uncertainties are given in table~\ref{table:localicd}.

While we have made an effort to provide good estimates for the
statistical uncertainties on $\xi$,  they likely underestimate the
real uncertainties, which may be dominated by systematic effects.
Indeed, several objects in table~\ref{table:localicd} have very low
formal, statistical errors (\ie, objects with $\delta(\xi) < 0.005$,
which are rounded to 0.00 in the table), which in reality are probably
underestimates.  Unaccounted systematic effects could include, for
example, inaccurate PSF--matching, image registration problems,
cosmetic effects in the case of the \uit\ images, etc.  Thus, we
caution that the high formal--S/N values of the flux measurements for
the ground--based $B$--band and \uit\ data probably correspond to an
underestimation of the uncertainties on $\xi$ as quoted in
table~\ref{table:localicd}.

\begin{figure}	
\plotone{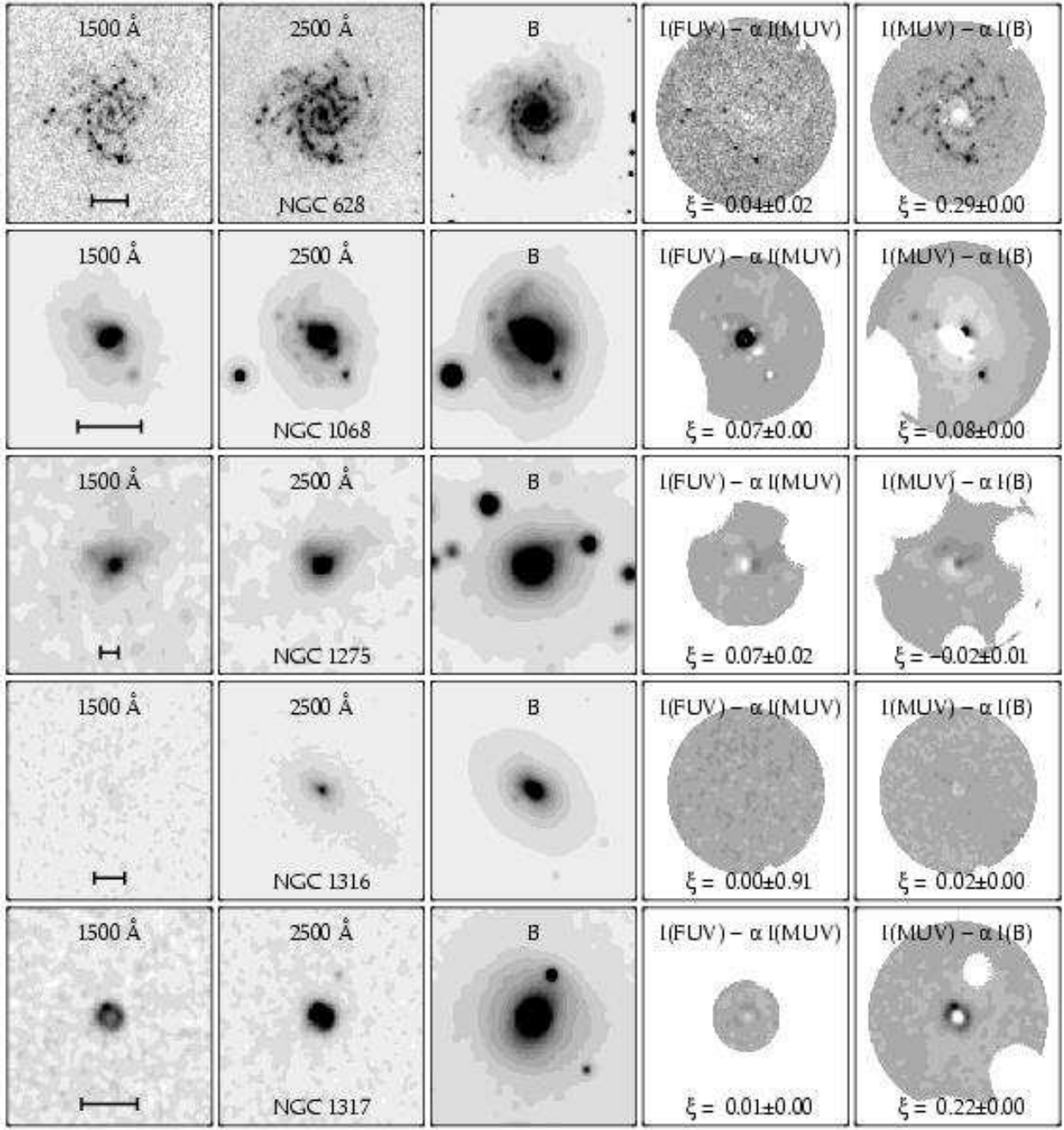}
\caption{Montage of color images for the \uit\ galaxy sample.  Each
row displays the \uit/FUV, \uit/MUV, and ground--based $B$--band image
(as labeled).  The last two columns show the residual map between
each band (\ie, $I_1 - \alpha I_2 - \beta$).  Note that this is
\textit{not} the same as the definition of $\xi$, which is the squared
sum of this image normalized by the squared sum of the object flux.
The corresponding $\xi$ value and derived uncertainty are indicated in
the bottom of these panels.  Horizontal bars in the first panel of
each row correspond to distances of 5~kpc at the rest-frame of the
galaxy. All images are displayed with north upward and east toward
the left.}\label{fig:montage}
\end{figure}

In figure~\ref{fig:montage}, we present a montage of images for each
galaxy in the \uit\ sample.  Each row of images in the figure shows
(from left--to--right) the \uit/FUV, \uit/MUV, $B$--band images, and the
residual images between these bandpasses after subtraction of the mean
flux ratio:  $f(\mathrm{FUV}) - \alpha\,f(\mathrm{MUV}) -
\beta$ and $f(\mathrm{MUV}) - \alpha\,f(B) - \beta$.  The
residual images illustrate galaxy features that exhibit regions of
strong variations in galaxy UV--to--optical colors relative to the
mean color value.  Dark/Bright regions correspond to regions 
that are bluer/redder than the mean color of the galaxy,
although note that these regions represent the flux difference (and
not the flux ratio, or {\it color}).

\begin{figure}
\plotone{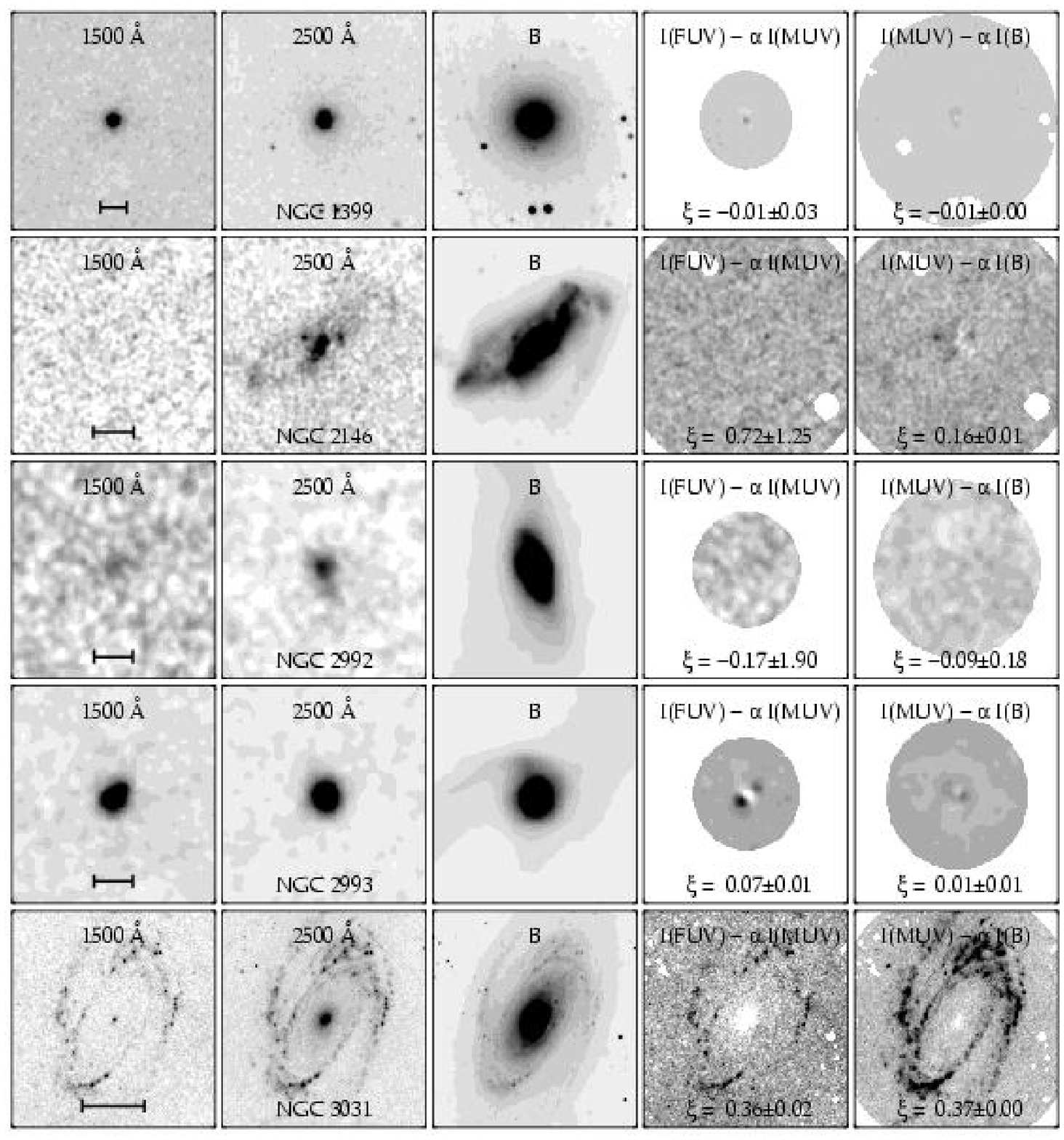} 
\figurenum{\ref{fig:montage}, cont}
\caption{}\label{fig:montage1}
\end{figure}

\begin{figure}
\plotone{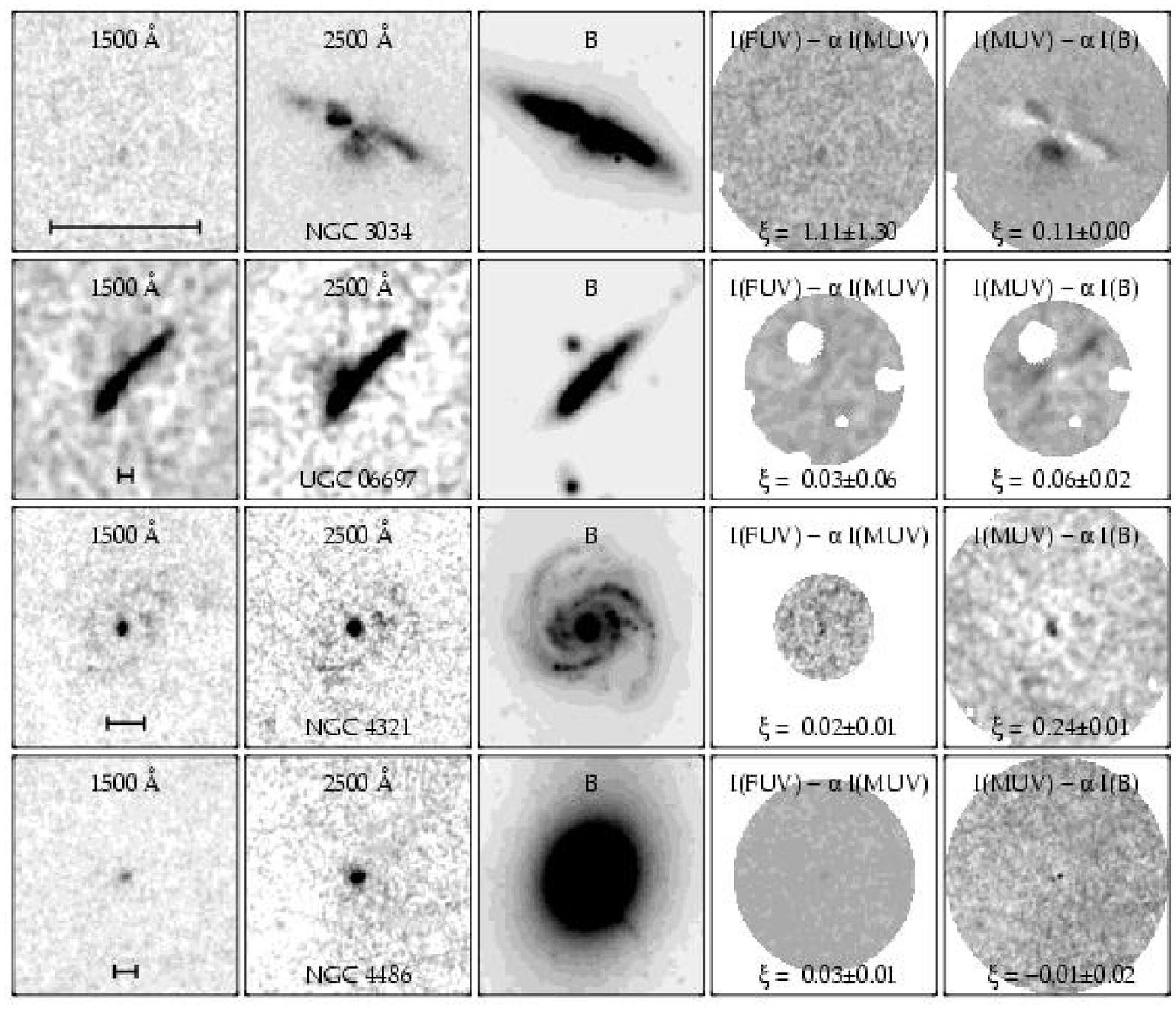}
\figurenum{\ref{fig:montage}, cont}
\caption{}\label{fig:montage2}
\end{figure}

Broad descriptions of the UV--to--optical morphologies of the galaxies
in our \uit\ sample have been presented by previous authors
\citep[see, \eg][and references therein]{mar01}, and we
refer the reader to those papers for further discussion on the general
morphological properties of these galaxies.  In the next section, we
discuss the distribution of internal UV--to--optical colors for the
galaxies in the \uit\ sample.

\subsection{Notes on Individual Objects}\label{section:notes}

\textit{NGC~628 (M74)}: NGC~628 is a near face--on galaxy with ``grand
design'' spiral arms.  In the UV, these spiral arms are traced by
bright, star--forming knots.  Diffuse stellar components constitute
the disk and bulge components, which dominate the $B$--band
morphology.  This galaxy has high $\xi(\mathrm{MUV},B)$, presumably
from differences in the UV/optical colors of the star--forming regions
and the more evolved stellar populations in the diffuse bulge/inner
disk.  There is little dispersion between the FUV and MUV internal
colors, which implies that the stellar populations that emit at these
wavelengths are largely co-spatial.

\textit{NGC~1068 (M77)}:  NGC~1068 is the nearest example of a type 2
Seyfert.  Differences in the UV/optical colors of the active
nucleus, bright star--forming knots, and  diffuse, extended disk
dominate the internal color dispersion (as shown in the residual color
image in figure~\ref{fig:montage}). To ascertain the contribution of
the AGN to the total internal color dispersion, we recomputed $\xi$
after masking out the galaxy center, and observed little change in the
resulting internal color dispersion.  Therefore,  the high internal
color dispersion value is  probably a result of patchy dust opacity
within the nuclear starburst (see also Neff \etal\ 1994).

\textit{NGC~1275 (Perseus A)}:  NGC~1275 is an early-type, cD galaxy
in the Perseus cluster.  \citet{con01} have interpreted
spectroscopic and morphological evidence to argue that NGC~1275 has
recently accreted one or more gas--rich cluster members, which is now
actively forming stars in the nucleus, and which has produced the
peculiar morphology.  The internal color dispersion between the MUV
and optical bands is low.  The residual image shows evidence for the
AGN--induced jet, as well as peculiar features associated with
star--forming regions.  The internal color dispersion between the FUV
and MUV is fairly significant, which apparently arises from a
combination of the AGN and star--forming regions.

\textit{NGC~1316 (Fornax A)}: The optical morphology of NGC~1316
displays prominent dust lanes (see also, Schweizer 1980).  No
measurable FUV--to--MUV internal color dispersion is apparent, but is
not excluded due to the low signal--to--noise ratio of the \uit/FUV
image.  The essentially nonexistent $\xi(\mathrm{MUV},B)$ value argues
that the stars contributing to the MUV and optical light are generally
co-spatial.

\textit{NGC~1317 (Fornax B)}:  The optical morphology of NGC~1317 is
dominated by the bulge/disk system.  The UV images indicate a
prominent nuclear ring of star formation (with radius $\sim 1$~kpc),
which is also observed in H$\alpha$ \citep{mar01}.   This ring
dominates the total UV emission and drives the internal color
dispersion and morphological $K$--correction between the UV and
optical images.

\textit{NGC~1399}:  The UV--to--optical morphologies of NGC~1399 are
broadly similar, and no measurable dispersion is seen in the internal
colors.  The UV emission probably stems from the exposed stellar cores
of evolved (old) low--mass stars (\ie, the ``UV-upturn'', see  Brown
\etal\ 2000), which is consistent from the very red colors, $m_{1500} -
B = 5.5$.  Regardless, the low $\xi$ values imply that the stars
dominating the flux from UV--to--optical wavelengths are well mixed.

\textit{NGC~2146}:  NGC~2146 is an inclined disk system, and exhibits
prominent dust features in its optical morphology. The MUV morphology
displays several knots in addition to a weak diffuse
component roughly co-spatial with the optical spiral-arm features.
The galaxy is only weakly detected at FUV wavelengths.  This, combined
with the high infrared emission from this galaxy, suggests that this
galaxy has undergone an obscured burst of star formation (Hutchings
\etal\ 1990).  The MUV--to--$B$-band internal color dispersion for
NGC~2146 is moderately high, resulting from apparently patchy dust
absorption, and supporting the obscured starburst hypothesis.  The
FUV--to--MUV internal color dispersion is immeasurable due to the low
signal--to--noise ratio of the \uit/FUV image.

\textit{NGC~2992/2993}: NGC~2992 and NGC~2993 represent an interacting
system, and exhibit extended tidal features in the optical images.
The MUV morphology of NGC~2992  shows faint, diffuse emission
coincident with the optical central bulge/disk and with low internal
color dispersion.  This galaxy is very faint in the FUV images.
NGC~2993 is well detected in both \uit/FUV and \uit/MUV images.  It is
also noteworthy that NGC~2993 has one of the bluest UV--to--optical
colors of the \uit\ sample.  This galaxy is apparently vigorously
forming stars and these stars dominate the UV portion of the galaxy's
SED.  The lack of internal color dispersion between the MUV and
optical images implies that the young stars also dominate the flux
emission at these wavelengths.  There is low (but significant)
internal color dispersion between the FUV and MUV images in NGC 2993,
which is centered on the nucleus and may be suggestive of a dust lane
or patchiness in the dust absorption.

\textit{NGC~3031 (M81)}:  NGC~3031 shows a dramatic transformation
between its optical and UV morphology.  Furthermore, this galaxy has
some of the largest $\xi$ values of the sample.  As illustrated by the
color--residual images for this galaxy in figure~\ref{fig:montage},
the large internal color dispersion values apparently result from
differences in the stellar populations that constitute the bulge,
disk, and star--forming spiral arms.  It is also notable that this
galaxy is the only sample member with high internal color dispersion
between the  FUV and MUV colors, \ie, $\xi(\mathrm{FUV},\mathrm{MUV})
> 0.2$.  This galaxy (along with NGC~3034) is the nearest galaxy in
the sample (at a distance of 3.4~Mpc; Freedman \etal\ 2001), and
therefore, one could postulate that the higher resolution available
for this galaxy somehow increases the internal color dispersion.
However, our simulations of NGC~3031 with poorer resolution indicate
that the resolution is not a dominant effect (see \S~5).  Therefore,
in order to account for the high internal color dispersion, the
(older) stellar populations of the bulge must contribute strongly to
the flux in MUV wavelengths (with very little accompanying FUV
emission), whereas the FUV flux predominantly stems from young
star--forming complexes in the spiral arms.  NGC~3031 beautifully
illustrates how differences in the composition of the stellar
populations that make up each galaxy component (\ie, the bulge, disk,
spiral arms) can generate large internal color dispersion.

\textit{NGC~3034 (M82)}: The optical morphology of NGC~3034 is complex
--- its inclination is nearly edge--on, and dust lanes clearly pervade
the UV and optical morphology, giving it  a patchy color distribution.
NGC~3034 is only marginally detected in the \uit/FUV image.  Intrinsic
FUV emission, which is presumably present due to the strong H$\alpha$
emission \citep{bel01}, is completely obscured by the large dust
opacity.  The patchy color distribution gives rise to significant
internal color dispersion between the MUV and $B$--band images.  The
residual color image in figure~\ref{fig:montage} illustrates
the regions of strong color dispersion, which appear to coincide with
the dust lanes in the optical and MUV images.  This illustrates an
example of internal color dispersion resulting from a varying amounts
of dust absorption.

\textit{UGC~06697}: UGC~06697 is well--detected in both \uit\ bands
with very blue UV--to--optical colors. The UV morphology is
broadly similar to the $B$--band image, even though it is nearly
edge--on (cf., NGC~3034).  There is little internal color dispersion
between the FUV and MUV bandpasses, which implies that the stellar
populations that emit at these wavelengths are largely co-spatial.
There is strong dispersion between the MUV and $B$--band internal
colors.  From the appearance of the residual image in
figure~\ref{fig:montage}, and the inclined nature of the system, this
dispersion is likely due to patchy dust obscuration.
 
\textit{NGC~4321 (M100)}:  The $B$--band image of NGC~4321 displays
grand spiral arms embedded in a large disk, with a small inner bulge.
The UV images have low signal--to--noise ratios.  A nuclear point
source dominates the \uit/FUV and MUV images, which is probably due to
nucleated star formation, but may result from a weak AGN (see Roberts,
Schurch, \& Warwick 2001).  Faint UV emission traces the spiral arms.
The internal color dispersion between the \uit/FUV and \uit/MUV bands
is inconclusive due to the low signal--to--noise ratios of the images.
NGC~4321 exhibits internal color dispersion between the \uit/MUV and
$B$--band images. This dispersion appears to result from color
differences between the nuclear region and the stellar populations of
the bulge and disk.

\textit{NGC~4486 (M87)}: NGC~4486 has the reddest UV--to--optical
colors in the \uit\ sample, and exhibits a very symmetric, smooth
morphology in all the images.  The MUV morphology is compact
(although some evidence for the nuclear jet is evident) and centered
on the nucleus, and is attributed to the exposed cores of
evolved stars.  There is also some indication for recent
star--formation, possibly resulting from a cooling flow in the cluster
\citep{mcn89}.  In a study  using deeper \uit/FUV images from the
\uitastro{2} mission, \citet{ohl98} conclude that only a few percent
of the FUV emission from NGC~4486 originates from an AGN or jet and that
most of the observed light is stellar in origin.   No measurable
internal color dispersion between the UV/optical colors is evident in
NGC~4486.


\section{Discussion\label{section:results}}

\begin{figure}
\epsscale{1.0}
\plotone{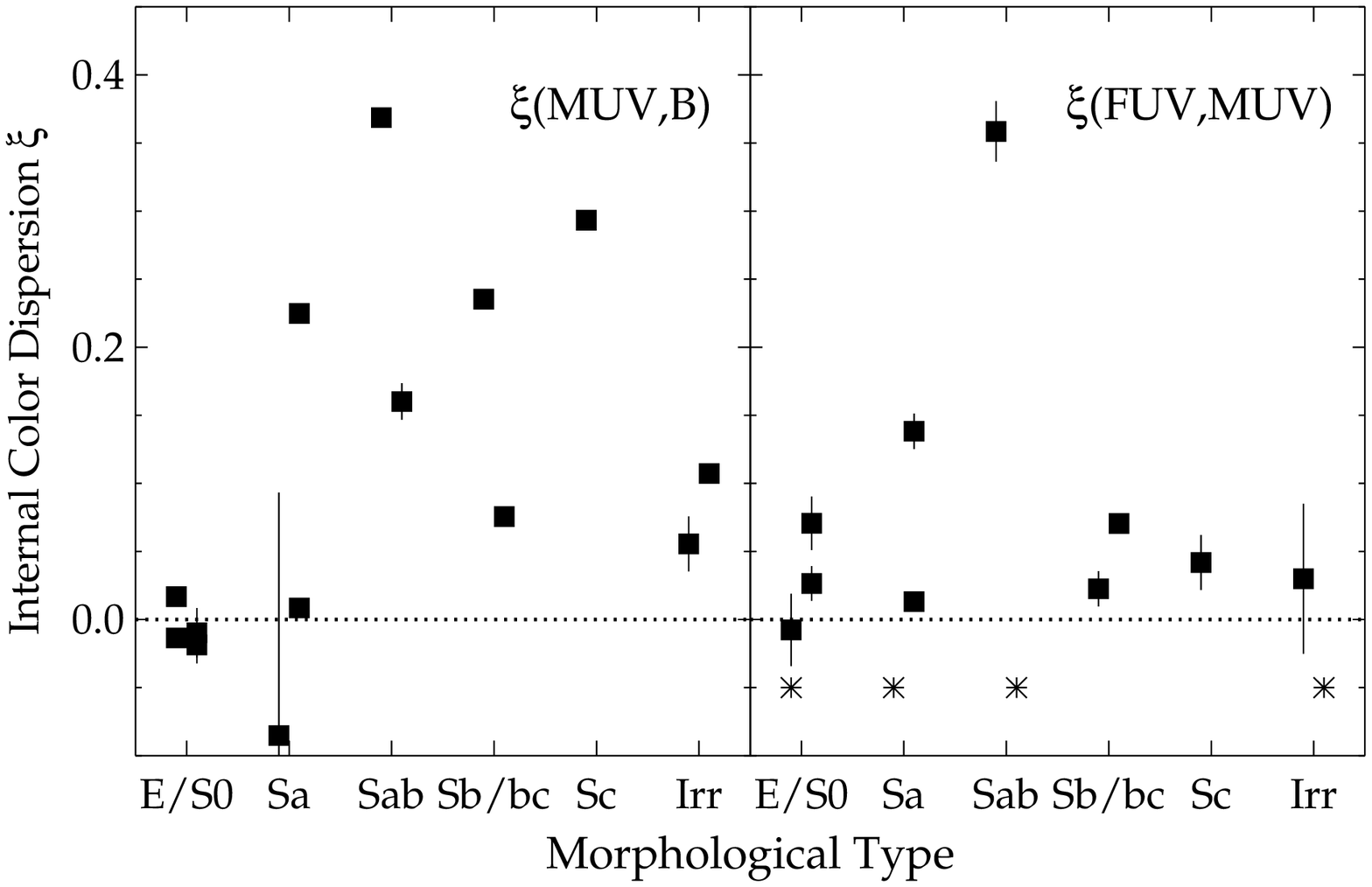}
\epsscale{1.0}
\caption{The derived internal color dispersion, $\xi$, as a function of
galaxy morphological type.  \textit{Left}: The panel shows the internal color
dispersion between the MUV-- and $B$--bands. \textit{Right}: The panel
shows the internal color dispersion between the FUV-- and MUV--bands.
Objects with $\mathrm{S/N} \leq 80$ in the \uit/FUV are indicated by
asterisks and defined to have $\xi(\mathrm{FUV},\mathrm{MUV}) \equiv
-0.05$ in the right hand panel. Note that points have been randomly
shifted by small amounts along the abscissa for clarity. }\label{fig:zetavtype}
\end{figure}

Because we expect the UV--optical internal color dispersion to be
sensitive to variations in the composition and configuration of the
underlying stellar population (and the strength of any AGN activity),
it is useful to compare the internal color dispersion with galaxy
morphology, which we show in figure~\ref{fig:zetavtype}.  The
MUV--optical internal color dispersion rises (left panel of
figure~\ref{fig:zetavtype}) from near zero for the earliest galaxy
types, peaks for the mid--to--late-type spirals and declines for
irregular galaxies.  This behavior is similar to the qualitative
trends observed between galaxies' morphological type (classified from
optical--wavelength images) and the strength of their morphological
$K$--corrections between UV and optical wavelengths, but here we have
measured this effect quantitatively.  The elliptical and lenticular
galaxies in the sample have low internal color dispersion between the
MUV--optical, \ie,  $\xi(\mathrm{MUV},B) < 0.02$, with a mean value,
$\langle \xi(\mathrm{MUV},B) \rangle \approx 0.0$, and small scatter,
$\sigma(\xi) = 0.02$.  Two of the Sa galaxies also show  low
$\xi(\mathrm{MUV},B)$ values (NGC~2992, NGC~2993), although the other
Sa galaxy in the sample (NGC~1317) has a high value,
$\xi(\mathrm{MUV},B) = 0.22$, which arises from the
star--forming, circumnuclear ring that dominates the UV morphologies
(see \S~\ref{section:notes}).  In contrast, the mid--type spirals
(types Sab--Sc) in the sample all have significant MUV--optical
internal color dispersion, with a mean value of
$\langle \xi(\mathrm{MUV},B) \rangle = 0.23$ and large scatter,
$\sigma(\xi) = 0.11$.  Lastly, the irregular galaxies in the sample
have moderate MUV--optical internal color dispersion, with a mean
value of $\langle \xi(\mathrm{MUV},B) \rangle \simeq 0.09$.  While this
is significant and is similar to the lowest values seen in the
mid--type spirals, it is less than the mean value seen in the
mid--type spirals.

Interestingly, the FUV--MUV internal color dispersion is much weaker
that that seen between the MUV--optical colors (contrast the left and
right panels of figure~\ref{fig:zetavtype}), and is roughly
independent of morphological type.  This suggests that the
morphological $K$--correction between the FUV--MUV bands is generally
small relative to that between the MUV--optical bands.  This result is
fairly unexpected, as the wavelengths of the chosen bandpasses span
similar baselines in logarithmic units, \ie, $\log (1500\,\AAA /
2500\,\AAA) \approx \log ( 2500\, \AAA / 4400\, \AAA)$.  Therefore,
the stars that dominate the FUV--to--MUV flux in the galaxies in this
sample are rather co-spatial compared to the configuration of the
stars that dominate at optical wavelengths.  The sole exception to the
above statements is NGC~3031 (M81), which shows very high FUV--MUV
internal color dispersion.  This high value apparently stems from the
fact that the bulge of NGC~3031 contributes a substantial portion of
the flux at MUV wavelengths, whereas the FUV emission is confined to
star--forming regions in the spiral arms (see \S~\ref{section:notes}).
Furthermore, this effect does not seem to be a consequence of
resolution, as we will discuss in more detail in \S~\ref{section:res}.
Therefore, while high FUV--MUV internal color dispersion values are
physically plausible, they seem to occur only for specific (rare?)
circumstances, and low $\xi$--values between FUV and MUV wavelengths
(\eg, small morphological $K$--corrections) are the norm.

\begin{figure}
\epsscale{1.0}
\plotone{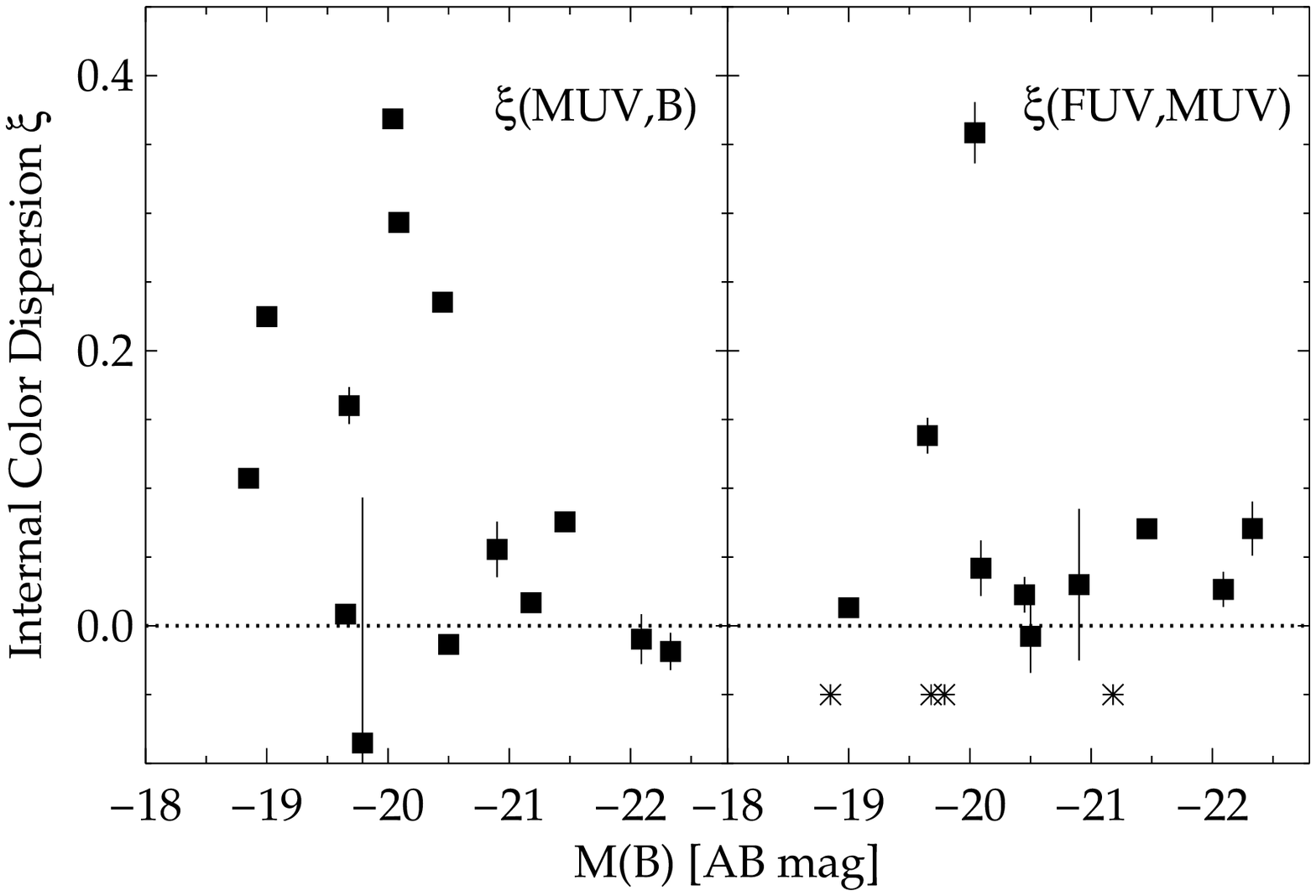}
\epsscale{1.0}
\caption{The internal color dispersion, $\xi$, versus the galaxy absolute
magnitude in the $B$--band (in units of AB magnitudes).  \textit{Left}:
The panel shows the color dispersion between the MUV-- and
$B$--bands.  \textit{Right}: The panel shows the internal color
dispersion between the FUV-- and MUV--bands.  Asterisks have the same
definition as in figure~\ref{fig:zetavtype}.}\label{fig:zetavbmag} \end{figure}

Because the mean absolute magnitude roughly correlates with
morphological type for local galaxies \citep[\eg,][]{rob94}, one can
study the dependencies between the internal color dispersion and
galaxy absolute magnitude, as illustrated in
figure~\ref{fig:zetavbmag}.  In general, objects with the highest
luminosities have low scatter in their derived $\xi(\mathrm{MUV},B)$
values, whereas at lower luminosities objects display greater scatter
in $\xi(\mathrm{MUV},B)$.  In contrast, there is no correlation
between absolute magnitude and $\xi(\mathrm{FUV},\mathrm{MUV})$.  This
is broadly consistent with the lack of dependency of the internal
color dispersion at these wavelengths and morphological type discussed
above.   Splitting the \uit\ sample into three luminosity sub--groups,
we find that for $M_B \le -21$, the dispersion in $\xi$ is
$\sigma(\xi(\mathrm{MUV},B)) = 0.042$ and
$\sigma(\xi(\mathrm{FUV},\mathrm{MUV})) = 0.026$.  In contrast, for
$-19 \ge M_B > -21$, we derive $\sigma(\xi(\mathrm{MUV},B)) = 0.155$
and $\sigma(\xi(\mathrm{FUV},\mathrm{MUV})) = 0.123$.  The increase in
the dispersion of $\xi(\mathrm{MUV},B)$ at this absolute magnitude is
due to the three mid--type spirals with large $\xi$ (NGC~3031,
NGC~628, NGC~4321).  For $\xi(\mathrm{FUV},\mathrm{MUV})$, the
increase is dominated solely by NGC~3031 (note, however that the value
for NGC~4321 is uncertain due to the low S/N of the \uit/FUV data).
For $M_B > -19$, the dispersion is $\sigma(\xi(\mathrm{MUV},B)) =
0.054$.  Thus, one conclusion from this analysis is that the most
luminous galaxies (in blue light) in the local universe have low
internal color dispersion between the UV and optical (and hence, the
weakest morphological $K$--corrections).  Less--luminous galaxies
($M_B \gsim -21$) exhibit examples of galaxies with high internal
color dispersion, and the scatter in these values on the whole is very
large relative to the more luminous galaxies.

\begin{figure}
\epsscale{1.0} 
\plotone{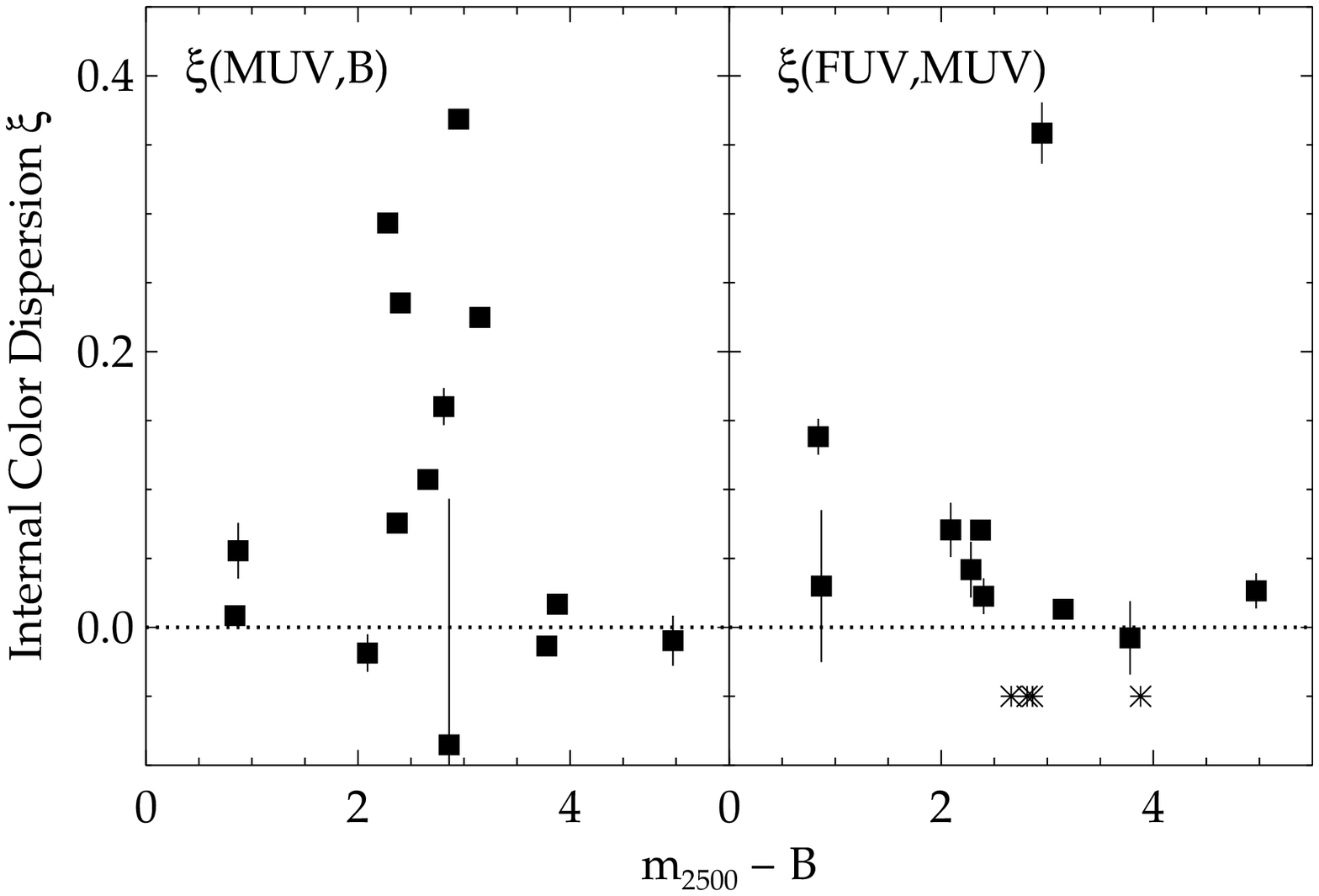} 
\epsscale{1.0}
\caption{The internal color dispersion, $\xi$, versus  MUV--to--$B$-band
total color, $m_{2500} - B$.  \textit{Left}: The 
panel shows the internal color dispersion between the MUV-- and
$B$--bands.  \textit{Right}: The panel shows the internal color
dispersion between the FUV-- and MUV--bands. Asterisks have the same
definition as in figure~\ref{fig:zetavtype}.}\label{fig:zetavnuvcolor}
\end{figure}

The dispersion of galaxy internal colors also exhibits trends with the
galaxy total color, as shown in figure~\ref{fig:zetavnuvcolor}.  There
is little relation between the FUV--MUV internal color dispersion and
UV--to--$B$-band color.  Galaxies with the reddest UV--to--$B$ colors,
\eg, $m_{2500} - B \gsim 3.5$, have very low MUV--$B$-band internal
color dispersion.  Thus, the stellar populations in these galaxies
that emit at these wavelengths have a very homogeneous distribution.
Due to the red colors and smooth (relaxed) morphologies of these
galaxies, any UV emission probably stems from the exposed cores of
evolved, late--type stars, which accounts for the co-spatial nature of
the UV and optical emission.  Similarly, the galaxies with bluest
colors, $m_{2500} - B \lsim 2$, have low--to--intermediate internal
color dispersion.  The light from galaxies with these colors is
dominated by young stellar populations (with ages less than a few
hundred Myr; in the absence of AGN activity).  Therefore, it seems
that when young stars dominate the total UV--to--$B$-band emission,
there is low intrinsic internal color dispersion, and any observed
internal color dispersion is likely the result of variable dust
absorption (although the effects of any such dust absorption are
limited such that they do not redden the galaxy total colors out of
this color range).  This seems to be the case for the latest
morphological types.  Note that this does not exclude the possibility
that older stars exist beneath the glare of the young stellar
populations, simply that any hypothetical old population does not
contribute substantially to the internal color dispersion or total
colors.

Galaxies in the sample with intermediate UV--to--optical colors (\ie,
$2 \lsim m_{2500} - B \lsim 3$) exhibit the largest range of internal
color dispersion values, which is illustrated by the ``swath'' of
$\xi(\mathrm{MUV},B)$ values that span $0 \lsim \xi \lsim 0.4$ in the
left panel of figure~\ref{fig:zetavnuvcolor}.  One can consider the
argument that these UV--to--optical colors may be in some sense a
requirement for the existence of high internal color dispersion.
Several possibilities then emerge.  The galaxies must have young
stellar populations that dominate the UV emission and are segregated
in some manner from a substantial evolved stellar population (\eg, a
disk/bulge dichotomy).  Otherwise, if young stars dominate the
UV--to--optical emission relative to the older stellar population,
then weak internal color dispersion is expected.  Similarly, weak
internal color dispersion is expected if the older stellar population
dominates the UV--to--optical emission (as is generally the case in
elliptical/lenticular galaxies; see previous paragraph).   Therefore,
it appears that the mix of young and old stellar populations --- with
integrated, intermediate $m_{2500} - B$ colors ---  is a condition
for high UV--optical internal color dispersion.

To exemplify this scenario, we used a simple model to investigate the
relationship between differences in galaxy stellar populations and the
inferred internal color dispersion.  We constructed a hypothetical
galaxy in which some fraction of resolution elements contain a young
stellar population and the remaining fraction contain an older stellar
population.   To model the UV--optical colors of the young and old
stellar populations, we used the latest version of the \citet{bru93}
stellar synthesis population models.  We modeled the colors of the
young population as having a Salpeter IMF and solar metallicity, and
having formed with constant star formation viewed at an age of
$10^8$~yr.  For the older stellar population, we considered two models,
both formed in an instantaneous burst with the same IMF and
metallicity as above, but viewed at ages $10^9$ and $10^{10}$~yr,
respectively.  These models produce UV--optical colors of, $m_{1500} -
B \simeq (0.9$, 2.1, 5.6) and $m_{2500} - B \simeq (0.8$, 2.6, 3.5),
for $t = (10^8$, $10^9$, $10^{10})$ yr, respectively.  We assigned these
colors to a model galaxy with uniform $B$--band surface brightness,
then vary the fraction of the galaxy dominated by the young stellar
population (where the remaining fraction of the galaxy is dominated by
the older stellar populations), and computed the internal color
dispersion values, $\xi(\mathrm{FUV,MUV})$ and $\xi(\mathrm{MUV},B)$.

\begin{figure}
\epsscale{1.0}
\plotone{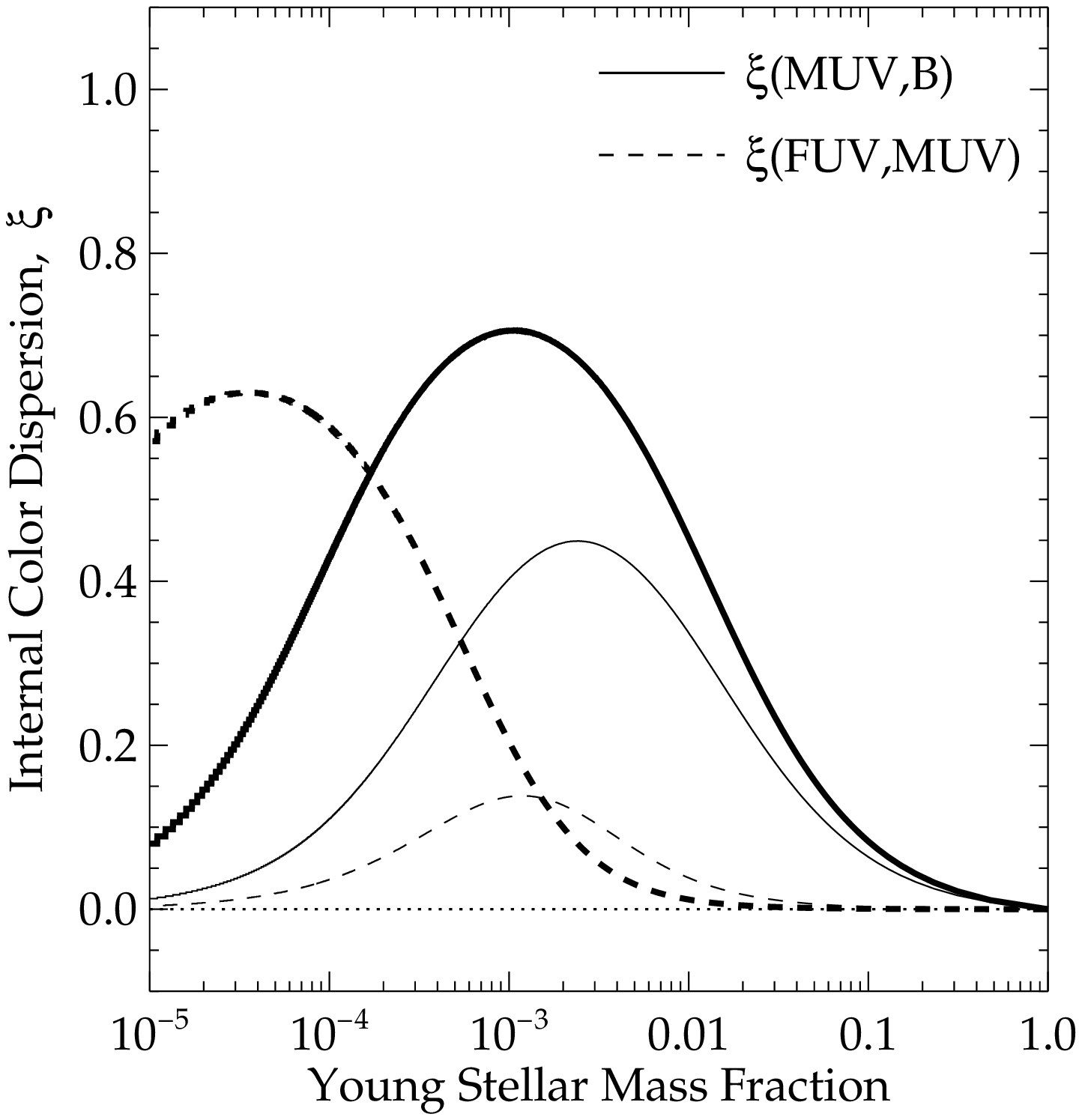}
\epsscale{1.0}
\caption{The internal color dispersion of simulated galaxies as a function
of the fraction of the total stellar mass in young stellar populations
(where the total stellar mass equals the sum of the mass in old
stellar populations and young stellar populations, see text).  For
each simulated galaxy, a fraction of the pixels is given the
UV--to--optical colors and stellar-mass--to--light ratios of a stellar
population formed with constant star formation and observed at an age
$10^8$~yr, with the remaining pixels given the colors and
stellar-mass--to--light ratios of a population formed in an
instantaneous burst and observed at an age $10^9$~yr
(\textit{thin} lines) or age $10^{10}$~yr (\textit{thick} lines).  The
curves show the internal color dispersion values between the FUV/MUV--
(\textit{dashed} lines), and MUV/$B$--bands (\textit{solid}
lines).}\label{fig:fracvxi}. \end{figure}

It is useful to parameterize how the UV--optical internal color
dispersion values for these galaxy models vary as a function of the
fraction of the total stellar mass made up by the young stellar
populations.  The mass--to--light ratios of the stellar populations
(plus stellar remnants) in the old and young stellar populations
defined above differ substantially:  $\mathcal{M}/L_B = (0.13$, 0.90,
10.3), for $t= (10^8$, $10^9$, $10^{10})$~yr, respectively.  In
figure~\ref{fig:fracvxi}, we show the relation between the internal
color dispersion as a function of the `young' stellar--population mass
fraction, where the total stellar mass equals the sum of the `young'
and `old' stellar--population mass.  Only a small fraction ($\lsim
10$\% by mass) of  nascent stars (relative to the total galaxy stellar
mass) is required to produce strong MUV/$B$-band internal color
dispersion.  At higher stellar--mass fractions ($\gsim 10$\%), the
young stellar populations dominate the MUV--optical flux emission,
with essentially no internal color dispersion.  Low internal color
dispersion is again observed for very low young stellar--population
mass fractions ($\lsim 0.01$\%), where the older stellar populations
dominate the MUV--optical emission.  Furthermore, strong
$\xi(\mathrm{FUV,MUV})$ is only observed in the case that young
stellar populations constitute minuscule fractions ($\lsim 0.1$\%) of
the total galaxy mass.  For higher mass fractions, the young stars
completely dominate the FUV--MUV emission, with no color dispersion.
At the low `young' stellar--mass--fraction end, small
$\xi(\mathrm{FUV,MUV})$ values are only seen in the case that the
young stellar populations constitute a tiny fraction of the total
stellar mass (\ie, $\lsim 0.1$\% for an old stellar population age of
$10^9$~yr, and $\lsim 0.001$\% for an old stellar population age of
$10^{10}$~yr).  These tiny fractions suggest that the internal
color dispersion should provide a useful constraint on  ``frostings''
of star formation in otherwise passively evolving elliptical galaxies
\citep[\eg,][]{tra00}.  We conclude that in the case of segregated
young and old stellar populations, the UV--optical internal color
dispersion constrains the relative stellar--mass proportions of these
populations.

These simulations are crude, and do not include the
effects of varying dust opacity, IMF, metallicity, or the possibility
of more than two stellar populations.  Obviously, any factor that
contributes to the relative difference in the colors throughout a
galaxy will contribute to the overall internal color dispersion.  The
results here, however, illustrate how the internal color dispersion
depends on the fraction of a galaxies' stellar mass in young stellar
populations.

An alternative scenario exists to account for the range of internal
color dispersion observed at intermediate $m_{2500} - B$ colors in
which the galaxies' internal colors result from a variable
distribution of dust absorption.  Dust is usually associated with
star--forming regions \citep[\eg][]{cal01}, and its general effect is
to redden UV--optical colors in dust--enshrouded regions, although the
exact amount depends crucially on the chosen extinction law.
Increasing the non-uniformity in the amount of dust absorption causes
galaxies to appear more ``patchy'' in these colors
\citep[\eg,][]{cal94,cal00,rom02}, which increases the internal color
dispersion.  At the same time, because young stellar populations often
form within dense dust clouds, one expects them to have intermediate
(redder) UV--optical colors, like those associated with the wide
scatter in $\xi(\mathrm{MUV},B)$ seen in
figure~\ref{fig:zetavnuvcolor}.  Severe dust extinction
($A_\mathrm{UV}$ of many magnitudes) fades the emission from galaxies
beyond the detection limit --- especially in the \uit/FUV images as is
arguably the case for NGC~3034 and NGC~2146.  Note, however, that both
of these galaxies show strong $\xi(\mathrm{MUV},B)$, which supports
the hypothesis that strong (and variable) dust extinction is
intermixed with star--forming populations.

\begin{figure}
\plotone{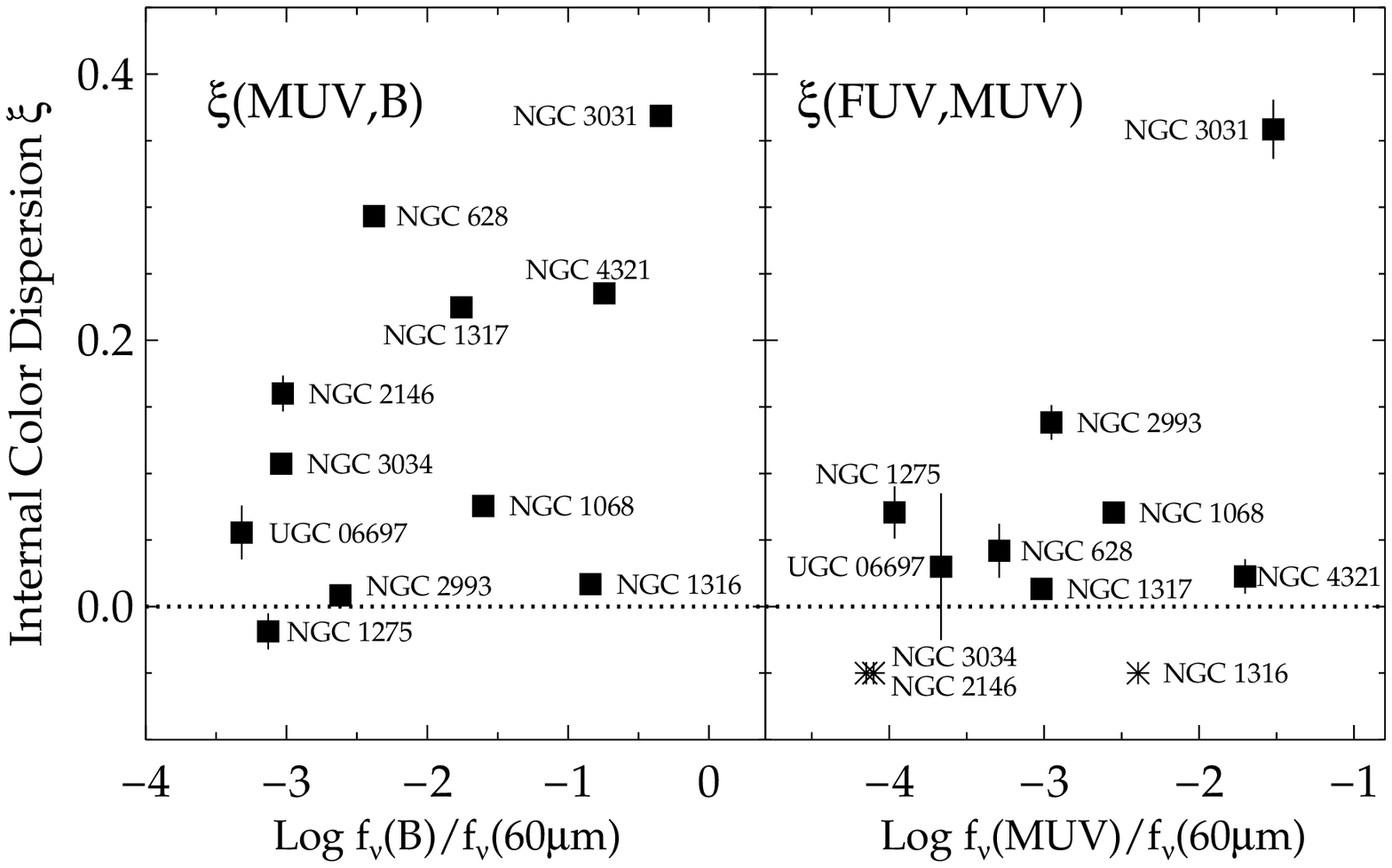}
\caption{Internal color dispersion versus UV/optical--to--FIR flux
ratio.  \textit{Left}: The panel depicts the internal color dispersion between
the MUV-- and $B$--bands as a function of $B$-band--to--FIR flux
ratio for those galaxies in the \uit\ sample and in the  \textit{IRAS}
catalogs.  \textit{Right}: The panel shows a similar result for the
internal color dispersion between the FUV and MUV as a function
MUV--to--FIR flux ratio.  The FIR flux density corresponds to that
measured by the \textit{IRAS} satellite through the 60~\micron\
channel.}\label{fig:zetavfir}
\end{figure}

If dust extinction is responsible for the internal color dispersion,
then there may be a relation between the internal color dispersion and
the reradiated emission from the dust, which should be detected in the
far--infrared (FIR).  To explore this possibility, we have compared
the derived internal color dispersion with the UV/optical--to--FIR
flux ratios for the galaxies in the \uit\ sample in
figure~\ref{fig:zetavfir}.  For the FIR flux emission, we used the
flux densities measured at 60~\micron\ from \textit{IRAS} (Rice \etal\
1988; Moshir \etal\ 1990; J.\ Knapp, 1994, priv.\ communication, see
the NASA/IPAC Extragalactic Database
(NED)\footnote{\texttt{http://ned.ipac.caltech.edu}}).  There is no
apparent relation between the $f_\nu(\mathrm{MUV})/f_\nu(60\micron)$
ratios and $\xi(\mathrm{FUV},\mathrm{MUV})$ values (see right panel of
figure~\ref{fig:zetavfir}). Interestingly, we observe a trend between
the $f_\nu(B)/f_\nu(60\micron)$ ratios and $\xi(\mathrm{MUV},B)$ (left
panel of figure~\ref{fig:zetavfir}) in that higher
$f_\nu(B)/f_\nu(60\micron)$ values correspond to an increase in the
scatter and upper envelope of $\xi$ values spanned by the data.   The
reason for this relation is not completely clear, but is probably a
result of how dust and stars are distributed in galaxies.  For
example, both NGC~3031 and NGC~4321 have high $f_\nu(B) /
f_\nu(60\micron)$ ratios, and correspondingly high
$\xi(\mathrm{MUV},B)$ values that probably result from differences in
the stellar populations that constitute the bulge/disk.
Alternatively, galaxies that appear to have more dust absorption
between the UV and optical images, \ie, NGC~3034, NGC~2146, and
UGC~06697, have lower $f_\nu(B) / f_\nu(60\micron)$ ratios.  Thus, variable
dust extinction apparently does induce UV--optical internal color
dispersion, but with smaller $\xi$ values relative to those produced
by variations in segregated stellar populations.  

This may imply that our internal color dispersion statistic is
more sensitive to differences in the underlying stellar populations
rather than the dust absorption.  For example, consider the effects of
adding a patchy dust screen on a galaxy forming a (single) young
stellar population.  Dust not only reddens the UV--optical colors, but
also fades the observed flux of the galaxy.  Because our internal color
dispersion statistic is normalized to the total galaxy flux (see
equation~\ref{eqn:zeta}), regions with the highest dust absorption
will contribute \textit{less} to $\xi$ relative to regions with less
dust.  Given the complex nature of dust (\ie, its geometry,
composition, configuration, temperature, etc.), in order to fully
parameterize the effects of dust on the UV--optical internal color
dispersion will require data for large galaxy samples that cover
UV--to--FIR wavelengths, which should be possible with the planned
surveys with the \textit{Space Infrared Telescope Facility}
(\textit{SIRTF})\footnote{see the \textit{SIRTF} Science Center
website: \texttt{http://sirtf.caltech.edu/SSC}.}.


\section{Application to Distant Galaxies}

The distribution of the internal color dispersion values for the \uit\
galaxy sample provides a benchmark for comparisons with high--redshift
galaxies.  In Paper~II, we compare these to observations of galaxies
in the \hdf\ using \hst/WFPC2 and NICMOS data out to $z\sim 3$.  In
this section, we consider the observational limitations of measuring
the internal color dispersion as a function of distance and resolution.

\subsection{Effects of Instrumental Resolution}\label{section:res}

\begin{figure}
\epsscale{0.8}
\plotone{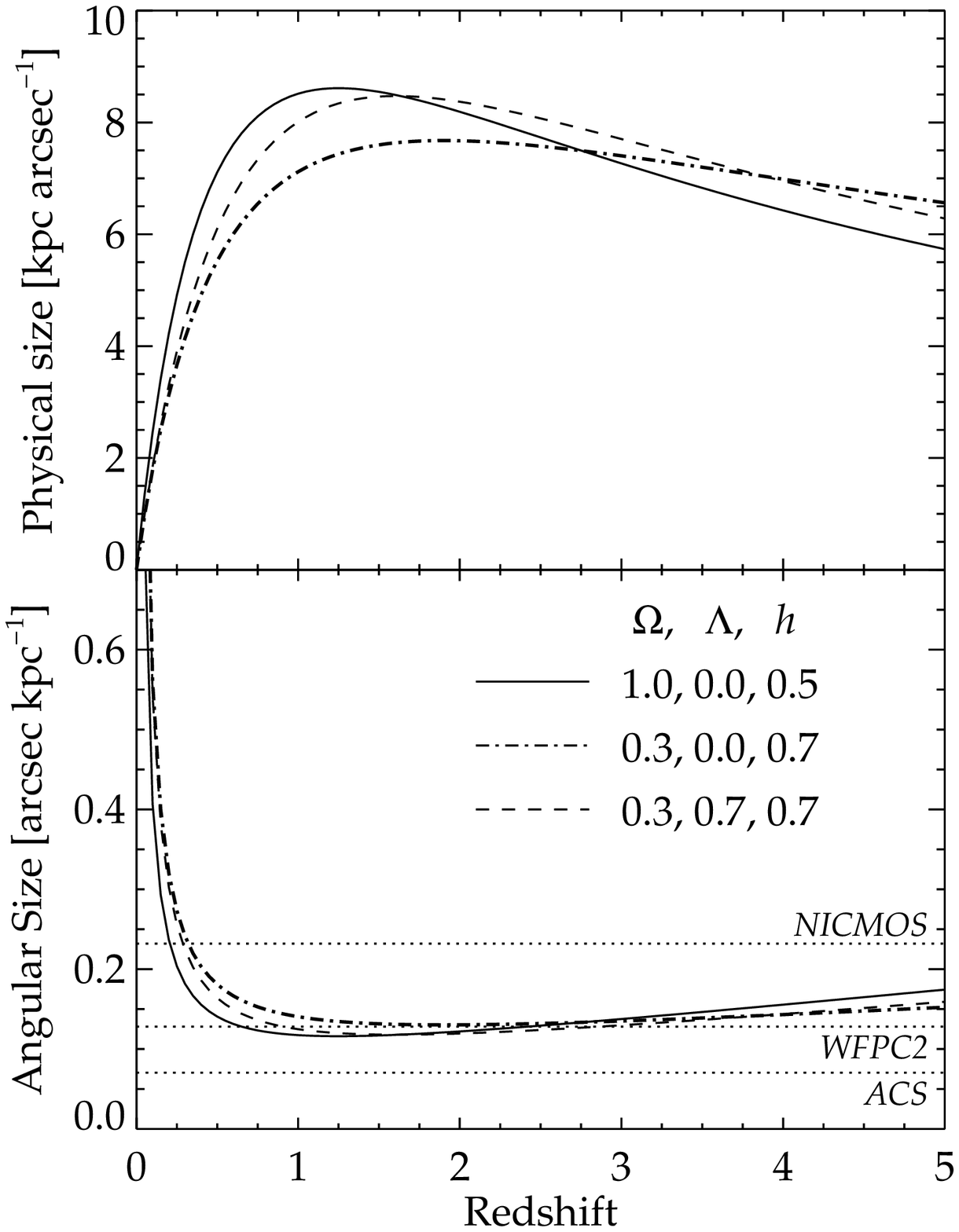}
\caption{The cosmological angular diameter distance as a function of
redshift for several fiducial world models.  The panels illustrate
the angular diameter distance, in units of the physical distance
corresponding to an apparent angular size (\textit{top} panel), and in
units of the angular size corresponding to a fixed physical distance
(\textit{bottom} panel).  The bottom panel also shows the nominal
resolving capacity of the current \hst\ instrumentation (as labeled;
note that the WFPC2 resolution corresponds to that of the WF chips,
and the NICMOS resolution is given for camera 3).
\label{fig:angdiam}} \epsscale{1.0} \end{figure}

One question is how our definition of the internal color dispersion varies
with the image physical resolution.  The images
in our \uit\ sample have a range of resolution depending on the
distance to the galaxy and the specific telescope/detector
characteristics.  It is clear that poor resolution will decrease the
measured internal color dispersion as intrinsic color variations
within a galaxy are ``smoothed over''.  For example, in the limiting
case where only one resolution element exists per galaxy, the integral
color dispersion would be zero.  Conversely, even a coeval and
statistically well--mixed stellar population that is resolved into
individual stars will have significant color dispersion.  Here, we
define the image resolution, $R$, as the number of resolution elements
(\ie, ``beams'') per kpc in the frame of the galaxy. For the \uit\
sample, these values range from $R \gsim 1-50$~beams~kpc$^{-1}$.  In
particular, resolution becomes important when considering observations
of distant (high--redshift) galaxies as one requires sufficient
resolution to measure quantitative morphological parameters.  For
example, the \hst\ observations that we analyze in Paper~II have a
nearly constant resolution, $R \sim 1-2$~beams~kpc$^{-1}$ for $z\simeq
0.7 - 3.5$ (due to the inflection in the angular diameter distance
about $z\simeq 1.6$ for the default flat, $\Lambda$CDM cosmology; see
figure~\ref{fig:angdiam}).

As an initial experiment, we considered the effects of limiting the
resolution for several fiducial galaxies that have moderate--to--high
internal color dispersion between the \uit/FUV, \uit/MUV, and $B$
passbands (NGC~3031, NGC~628, UGC~06697).  We convolved the galaxies
to mimic the effects of  larger ``beams'' to simulate observations of
each galaxy with decreasing resolution of the physical distance in the
frame of each galaxy.

\begin{figure}
\plotone{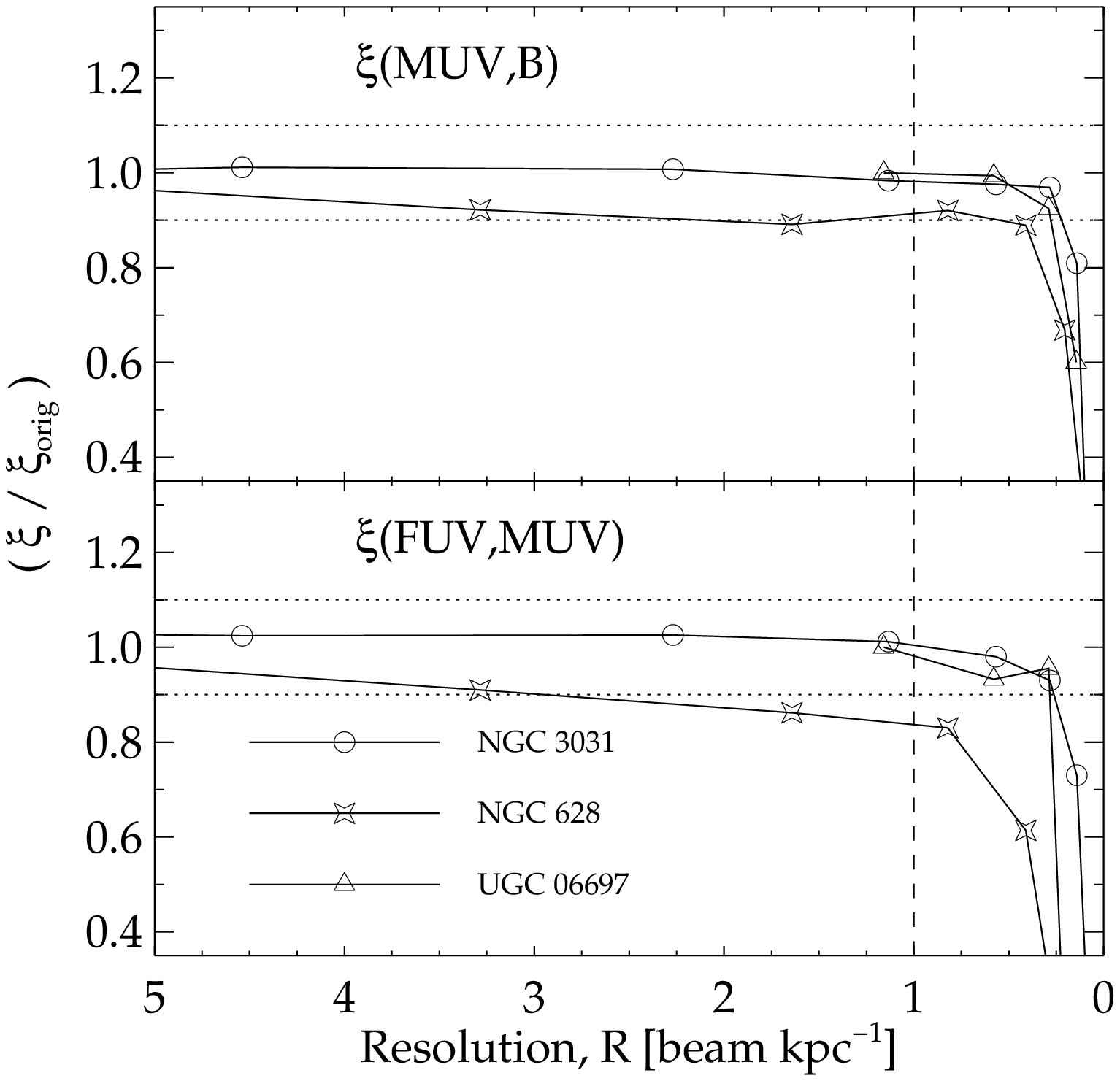}
\caption{The change in the derived internal color dispersion as a
function of image resolution.  Each curve corresponds to a fiducial
galaxy (listed in the plot inset) as a function of image resolution,
$R$, in units of beams per kpc.  The ordinate shows the change in the
internal color dispersion relative to the original value. In general,
changing the image resolution has a small effect on the measured
internal color dispersion for $R \gsim 0.5$~beams~kpc$^{-1}$.  Below
this value, any intrinsic color dispersion is suppressed by the poorer
resolution of the image.}\label{fig:rebinsim}
\end{figure}

In figure~\ref{fig:rebinsim}, we show the derived internal color
dispersion values (relative to the values for their original image
resolution) as a function of resolution for NGC~3031, NGC~628, and
UGC~06697.   Note that only the resolution has changed in these
simulations, \ie, the galaxy fluxes have been preserved (with no
losses due to cosmological surface brightness, for example).  In these
simulations, the internal color dispersion varies by $\lsim 10$\% down
to $R \gsim 0.5$~beam~kpc$^{-1}$.  Therefore, $\xi$ is a robust
measure of internal color dispersion (and thus the morphological
$K$--correction) down to this limiting resolution.  As a further test,
we compared the internal color dispersion values and image resolution
for the \uit\ galaxy sample.   Although the images span a large
resolution range, we find no relationship between the two.  Therefore,
reducing  the resolution of the \uit\ galaxies to $R \sim 0.5$~beam
kpc$^{-1}$ does not strongly impact the results presented here.

The resolution limit for measuring reliable internal color dispersion
values may depend on the intrinsic galaxy size.  Although giant spiral
galaxies commonly show significant internal structure (e.g., NGC~3031,
NGC 628), this is not the case for all large galaxies (e.g.,
ellipitcals, NGC~4486; and lenticulars, NGC~1316).  Similarly, our
\uit\ sample is lacking small, dwarf galaxies that could require high
resolution to measure any internal color dispersion.    There is
evidence that the sizes of star--forming complexes in spiral,
irregular, and dwarf galaxies scale approximately as the square root
the galaxy $B$--band luminosity \citep{elm94,elm96}.  The sizes of
star--forming complexes in brighter galaxies ($-15 \gsim M_B \gsim
-23$) are $\sim 0.1-3$~kpc, and are comparable to the physical length
of the limiting resolution for the internal color dispersion.  Because
the internal color dispersion between the \uit\ bands and $B$--band is
sensitive to heterogeneities in the star--forming regions of a galaxy,
it may be the case that robust measures of $\xi$ are possible given
that the image resolution is less than the sizes of the
star--formation complexes within an individual galaxy.  There is
insufficient \uit\ data available to test this paradigm over a
complete range of galaxy types, sizes, and star--formation properties,
but such studies may be possible using MUV--optical imaging of nearby
galaxies from the \hst\ \citep[\eg,][]{win02}.

What resolution is required then to reliably measure the internal
color dispersion of high--redshift galaxies?  As we mentioned
previously, the cosmological angular--diameter distance (the ratio of
the object's distance to its apparent size on the sky) exhibits a
well--known inflection around $z\sim 1 - 2$ (depending on the exact
cosmological parameters), as illustrated in figure~\ref{fig:angdiam}.
For the default cosmology used here ($\Omega_m=0.3$, $\Lambda=0.7$,
$h=0.7$), the maximum in the angular--diameter distance occurs near
$z\simeq 1.6$, with $D_A \simeq 8.5$~kpc~arcsec$^{-1}$.  For the
requirement that the image resolution exceed $R \gsim
0.5$~beam~kpc$^{-1}$ (see above), this corresponds to $\sim 0\farcs
1$~beam, which is generally only available from space--based
observatories, but also from ground--based telescopes with
adaptive--optics and interferometric techniques.  The \hst/ACS, WFPC2,
and NICMOS (camera 3) have beam sizes of $\mathrm{FWHM} \simeq
0\farcs07$, $0\farcs13$, and $0\farcs23$,\footnote{Because the \hst\
cameras are undersampled, the pixel scale dominates the beam size.
The values quoted here correspond to the FWHM measured from artificial
PSFs constructed using the TinyTim software provided by J.\ Krist
(private communication, see also
\texttt{http://www.stsci.edu/software/tinytim/}) at the pixel scale of
the \hst\ detectors.  Note that there are several techniques available
to reconstruct the image resolution in dithered, undersampled images
\citep[\eg,][]{fru02,lau99}, which offer improved image resolution
over the values presented here.} respectively, which are sufficient
for quantitative studies of the galaxies internal colors, and greater
than needed in the case of large multi--component galaxies like
NGC~3031 (see also the discussion in Conselice \etal\ 2000a).  Note,
however, that even the \hst\ resolution is nearly at the limit for
quantitative morphology for high--redshift galaxies, which argues that
telescopes with even higher resolving power would greatly improve our
ability to measure morphological properties of galaxies in the distant
universe.

\subsection{The Internal Color Dispersion of High--Redshift Galaxies}

Cosmological effects cause the surface brightness of galaxies to fade
strongly with redshift, the bolometric surface brightness decreases as
$\propto (1+z)^4$.  Many authors have argued that this complicates
interpretations of high--redshift galaxies as one is biased toward
high--surface brightness regions and may be missing a substantial
fraction of the total luminosity from galaxies \citep[see, for
example,][and references therein]{bah90,bur02}.  Indeed, comparing the
(artificially) redshifted appearances of local galaxies to \hst\
images illustrates the consequences of cosmological surface
brightness, especially when only rest--frame UV wavelengths are
considered \citep[see][]{gia95,kuc01}.

We have investigated the appearance of the \uit\ galaxies if viewed at
high--redshift in the deep \hst/ WFPC2 and NICMOS images for the \hdf\
\citep{wil96,dic02}.  We selected four galaxies from the \uit\ sample
with significant internal color dispersion in order to gauge how these
are affected if viewed at high redshift: NGC~3031, NGC~628, NGC~1068,
UGC~06697.  We have simulated their appearances at $z = 0.5$, 1.0,
1.5, 2.0, 2.5, and 3.0, at the pixel scale of the \hst\ \hdf\ images
($\simeq 0\farcs04$~pix$^{-1}$).  We first convolved these images with
a kernel such that the final simulated images match the  NICMOS Camera
3 PSF for the F160W filter at each redshift (in our analysis of the
NICMOS \hdf\ galaxies all images have been convolved to match the NIC3
F160W PSF, see Paper~II).  We then resampled each image and rescaled
the image pixel intensities to account for the change in distance
using the detailed prescription described in the Appendix.  For the
simulated images at $z=0.5$, 1.0, 1.5, we added noise from the
WFPC2/F300W, WFPC2/F606W, and NIC3/F110W data to the \uit/FUV,
\uit/MUV, and $B$--band images, respectively.  For $z=2.0$, 2.5, 3.5,
we instead add noise from the WFPC2/F606W, NIC3/F110W, and NIC3/F160W
images, as these bandpasses better correspond to the rest--frame
\uit/FUV, \uit/MUV, and $B$--band for these redshifts.

\begin{figure}
\epsscale{1.1}
\vbox{\plottwo{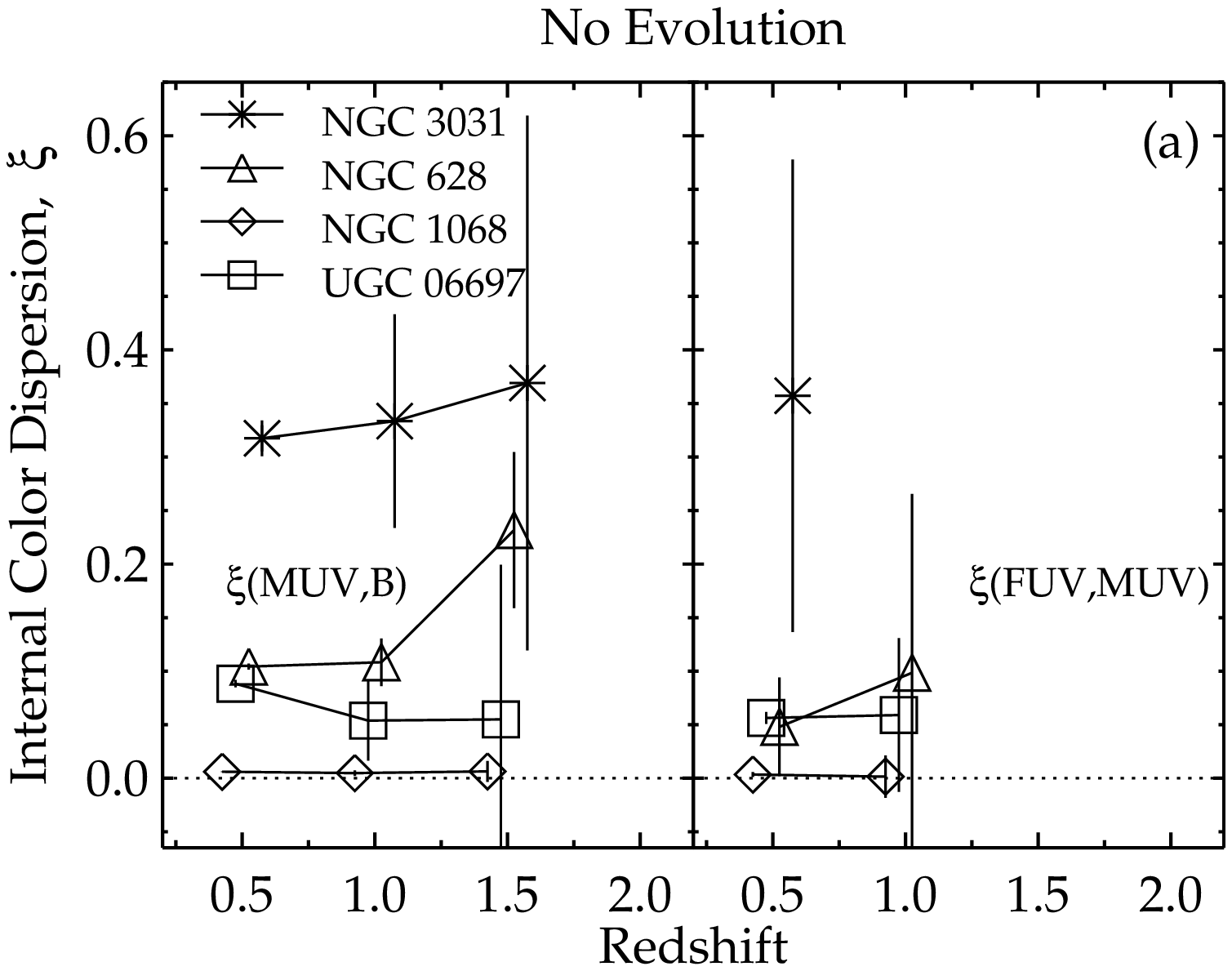}{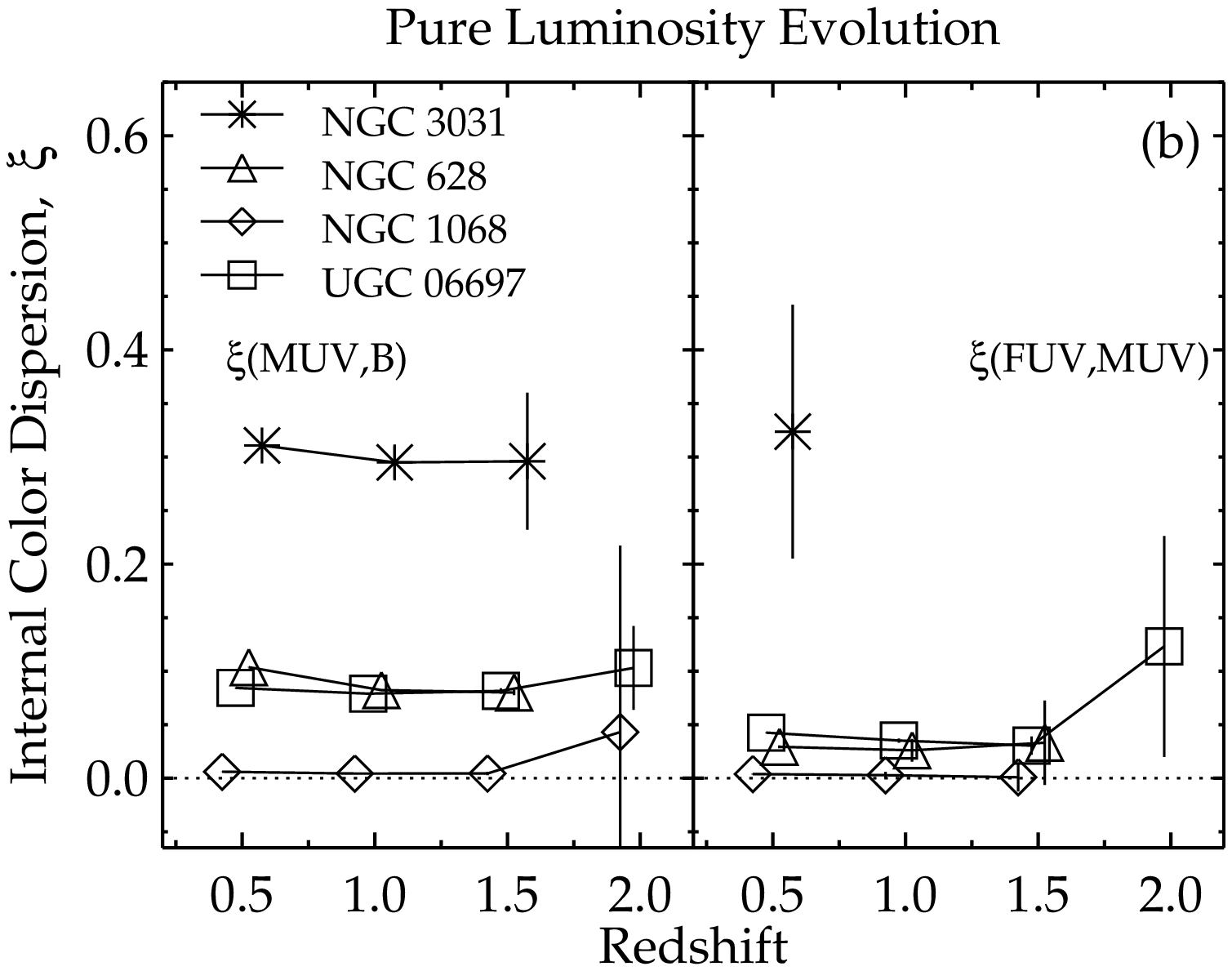}}
\vspace{0.25in}
\vbox{\plottwo{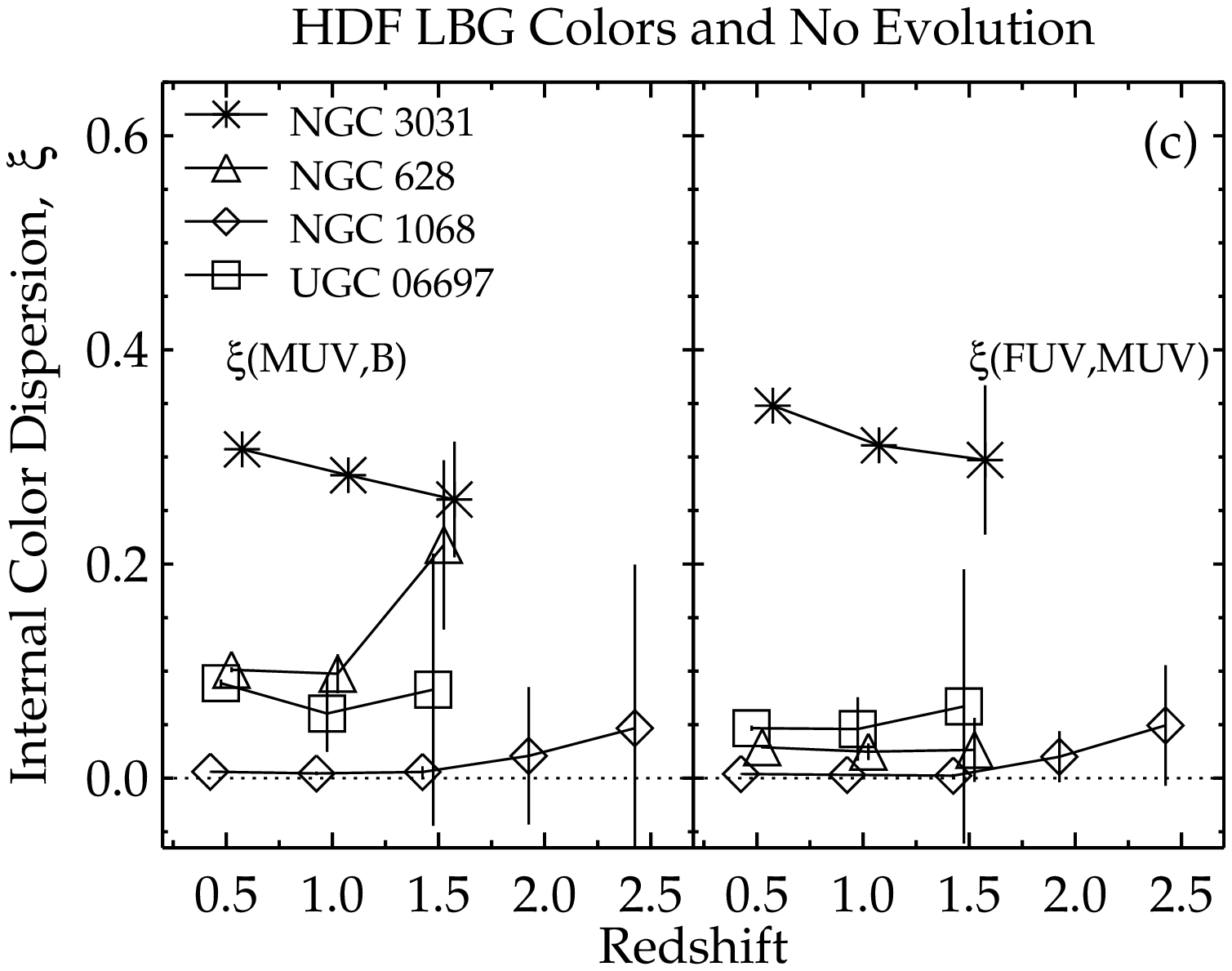}{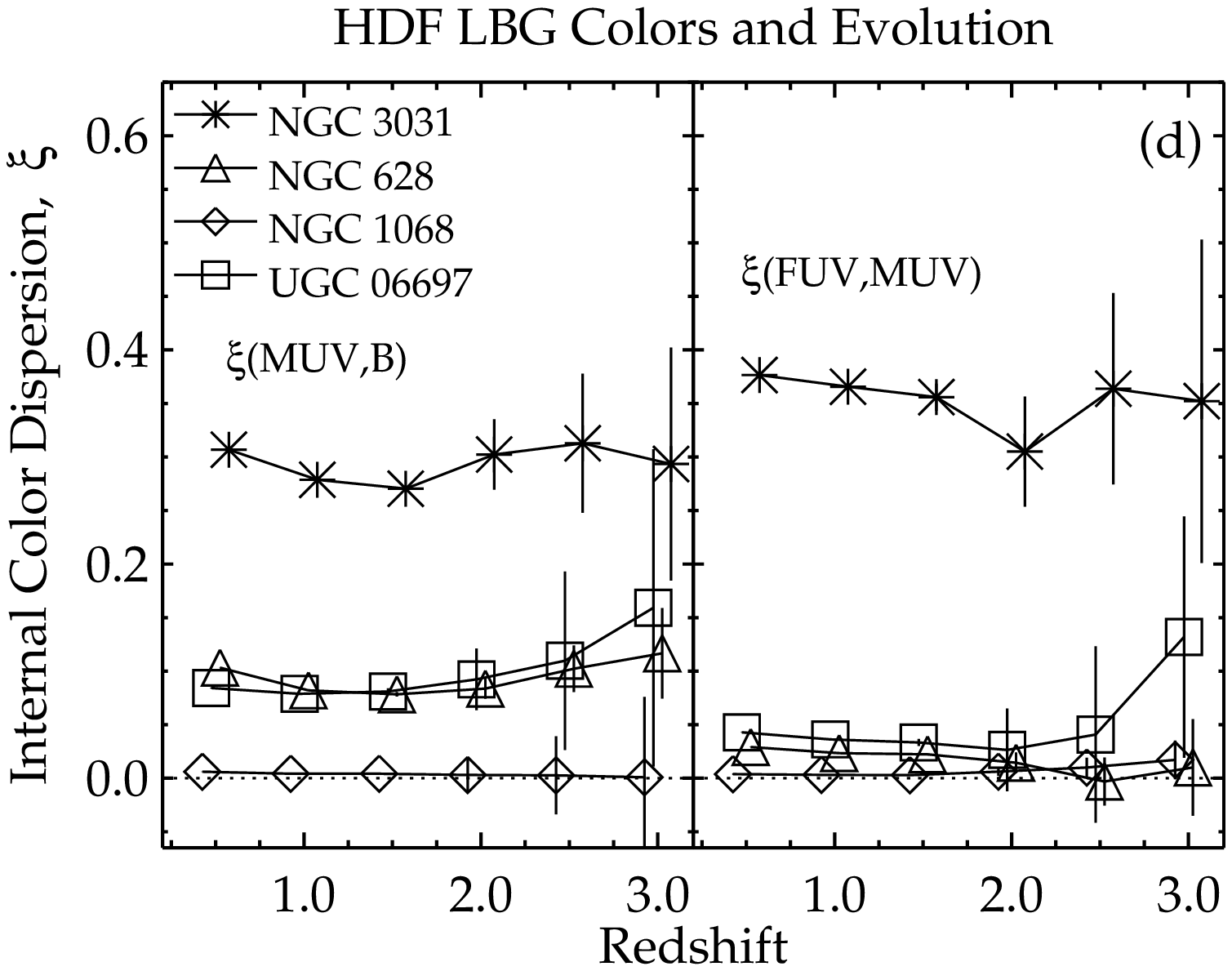}}
\caption{The internal color dispersion for four fiducial galaxies
(NGC~3031, NGC~628, NGC~1068, UGC 06697, as labeled) that have been
simulated at higher redshifts and inserted in \hst\ data for the \hdf\
(see text).  The four different panels (a--d) correspond to various
assumptions about the luminosity and color evolution of the galaxies.
Each panel shows $\xi(\mathrm{FUV},\mathrm{MUV})$ (\textit{left}
sub-panels) and $\xi(\mathrm{MUV},B)$ (\textit{right} sub-panels).
Panel (a) shows $\xi$ under the assumption of no evolution from
$z\approx 0$ to higher redshift.  Panel (b) shows results for
incorporating luminosity evolution (parameterized by boosting the
fluxes by $+1$~mag for NGC~3031 and NGC~1068, and by $+2$~mag for
NGC~628 and UGC~06697) while keeping the UV--$B$-band colors fixed.
Panel (c) illustrates the effects of no luminosity evolution, but
adjusting the galaxy colors from their present--day values ($z\approx
0$) to the approximate median (rest--frame) $m_{1500} - m_{2500}$ and
$m_{2500} - B$ colors observed in the \hdf\ LBGs (see text).  Panel
(d) shows the resulting internal color dispersions allowing for both
luminosity evolution [as in panel (b)] and color evolution [as in
panel (c)].  Note that points for each galaxy are not shown if the
simulation has insufficient S/N.\label{fig:rebinhdf}} \epsscale{1.0}
\end{figure}

Panel (a) of figure~\ref{fig:rebinhdf} shows the measured internal
color dispersion between the (rest--frame) \uit/FUV--MUV, and
\uit/MUV--$B$--bands as a function of redshift under the assumption of
no evolution (\ie, the luminosities and colors of the galaxies are
unchanged from their $z\approx 0$ values).  One salient conclusion
from this plot is that the internal color dispersion for these
galaxies between either set of bandpasses is immeasurable for $z\gsim
1-1.5$.  This roughly corresponds with the point where the
(rest--frame) UV fluxes diminish to $\mathrm{S/N} \lsim 80$, which our
previous simulations (\S\ref{section:icd}) indicate the internal color
dispersion is generally unreliable.  The galaxies are detected in the
rest--frame $B$--band data to $z\gsim 2$ (see below), but their
rest--frame UV flux emission at high redshifts (with no evolution) is
too faint for quantitative morphological studies.

In panel (b) of figure~\ref{fig:rebinhdf}, we show the derived
internal color dispersion values for the same galaxies as above, but
here we have increased the galaxies' luminosities while keeping their
UV--optical colors fixed at their present values.  This plot shows
that for NGC~628 ($\Delta m = +2$~mag), NGC~1068 ($\Delta m =
+1$~mag), and UGC~06697 ($\Delta m = +2$~mag), the dispersion in the
internal color is consistently robust to $z\gsim 2$.  These galaxies
have blue UV colors and thus the rest--frame UV images have sufficient
signal--to--noise with which to measure the internal colors, at least
to $z\sim 2.5$.  NGC~3031 ($\Delta B = +1$~mag), with $m_{1500} -
m_{2500} = 1.5$ and $m_{2500} - B = 3.0$ is relative red, and even
with the increase in luminosity, both $\xi(\mathrm{FUV},\mathrm{MUV})$
and $\xi(\mathrm{MUV},B)$ remain poorly measured for $z \gsim 2$.

High redshift galaxies are not only more luminous than present--day
counterparts, but they also have bluer rest--frame colors, which is
especially true for $z\gsim 1$ (\eg, Dickinson 1999;
Dickinson \etal\ 2003a).  To test the effects of color evolution on
$\xi$, we have scaled the observed UV--to--optical colors of the four
fiducial galaxies from the \uit\ sample to match observations at high
redshifts.  The Lyman--break galaxies (LBGs) in the \hdf, studied by
\citet{pap01} have observed median colors $(\wfv - \nicj)_\mathrm{med}
= 0.265$ and $(\nicj - \nich)_\mathrm{med} = 0.464$, which
approximately convert to $m_{1500} - m_{2500}$ and $m_{2500} - B$ in
the rest--frame of the galaxies.  Converting the UV--optical colors of
the \uit--sample galaxies to the these colors provides a useful
evolution scenario because high--redshift galaxies are empirically
\textit{observed} to have these colors.  Panel (c) of
figure~\ref{fig:rebinhdf} shows the measured internal color dispersion
as a function of redshift for the case of pure color evolution, with
no luminosity evolution (\ie, the $B$--band luminosities are unchanged
from the $z\approx 0$ values).  For this case, the internal color
dispersion in these galaxies is robust to $z\sim 1.5$ --- even for
$\xi(\mathrm{FUV},\mathrm{MUV})$ for NGC~3031, which now has high
enough S/N in the rest--frame UV images from which to accurately
derive $\xi$.

Combining both the luminosity and color evolution provides consistent
measures of the internal color dispersion for all the galaxies
simulated here out to $z \sim 2.5 - 3$.  In panel (d) of
figure~\ref{fig:rebinhdf}, we show the measured $\xi$ values for the
simulated \uit--sample galaxies, and including luminosity evolution
(by the amounts used in panel (b)) and color evolution used to match
the \hdf/LBG colors (as in panel (c)).  Note that we have implicitly
assumed no morphological evolution in these galaxies.  We
conclude that under these evolutionary scenarios with simple changes
in the color and luminosity, the intrinsic internal color dispersion
of each of these galaxies is robustly measured even for $z\sim 3$.
This seems to suggest that if the \uit--sample galaxies were observed
at $z\approx 0$ with the luminosities and UV--optical colors that are
characteristic of high--redshift galaxy populations, then their
UV--optical internal color dispersion values would be measurable in
deep \hst--like observations.

Inspecting the simulated images for each galaxy produces some insights
into the observed morphological and internal color changes as a
function of redshift. The internal color dispersion in NGC~1068 is
suppressed as the galaxy is transformed under the \hdf\ conditions.
The reduction in signal apparently results from the convolution with
the NICMOS PSF and adding the backgrounds of the \hdf\ data.  Thus, the
internal colors of distant galaxies may have some bias against
systems in which the internal color dispersion is dominated by a
combination of very a nucleated starburst and AGN activity.

\begin{figure}
\plotone{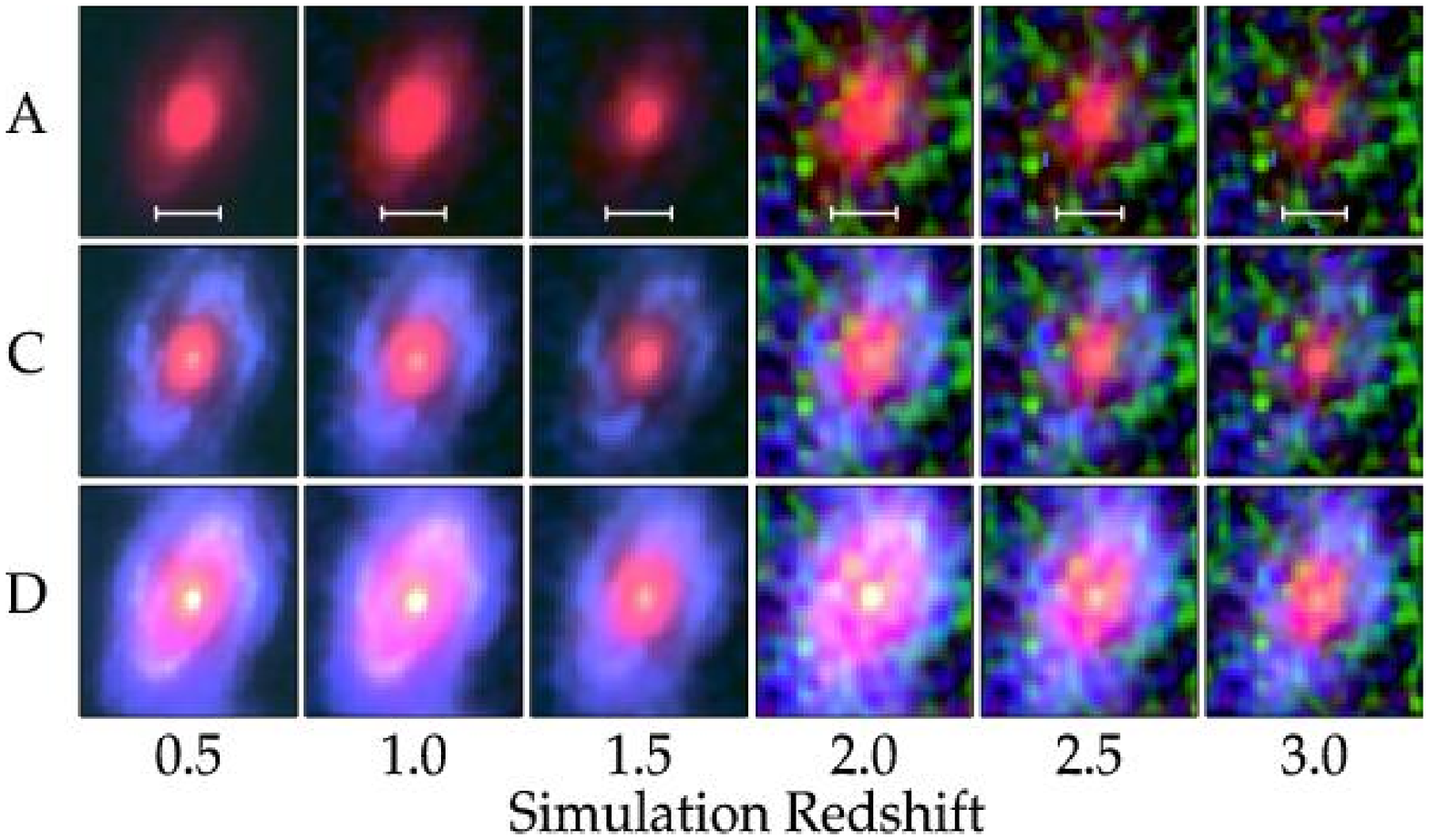}
\caption{Montage of color images illustrating the appearance NGC~3031
(M81) as it would appear in the WFPC2/NICMOS \hdf\ data for increasing
redshift (as indicated on the abscissa).  Colors correspond to
(rest--frame) FUV (blue), MUV (green), and $B$--band (red).  The image
has been convolved with the NICMOS camera 3 F160W PSF, rescaled to the
NICMOS pixel scale at each redshift, and noise has been added to the
\uit/FUV, \uit/MUV, and $B$--band images from the \wfu, \wfv, \nicj\
images, respectively, for $0.5 \leq z \leq 1.5$ and \wfv, \nicj, \nich\
images, respectively, for $2.0 \leq z \leq 3.0$ , respectively (see
text).  The top row (Row A) illustrates the appearance of the galaxy
with no evolution.  The middle row (Row C) shows the galaxy with no
luminosity evolution but with color evolution such that the
FUV--to--optical colors correspond to the median values of the \hdf\
LBG sample.  The bottom row (Row D) illustrates the same color
evolution as in row C, but the luminosities have also been increased
by one magnitude in each bandpass. Note that the contrast has been
adjusted slightly in each image to best illustrate the features.  The
horizontal bar in each panel of the top row corresponds to a distance
of 5~kpc in the rest frame of the galaxy at each
redshift.}\label{fig:zsimngc3031} \end{figure}

\begin{figure}
\epsscale{1.0}
\plotone{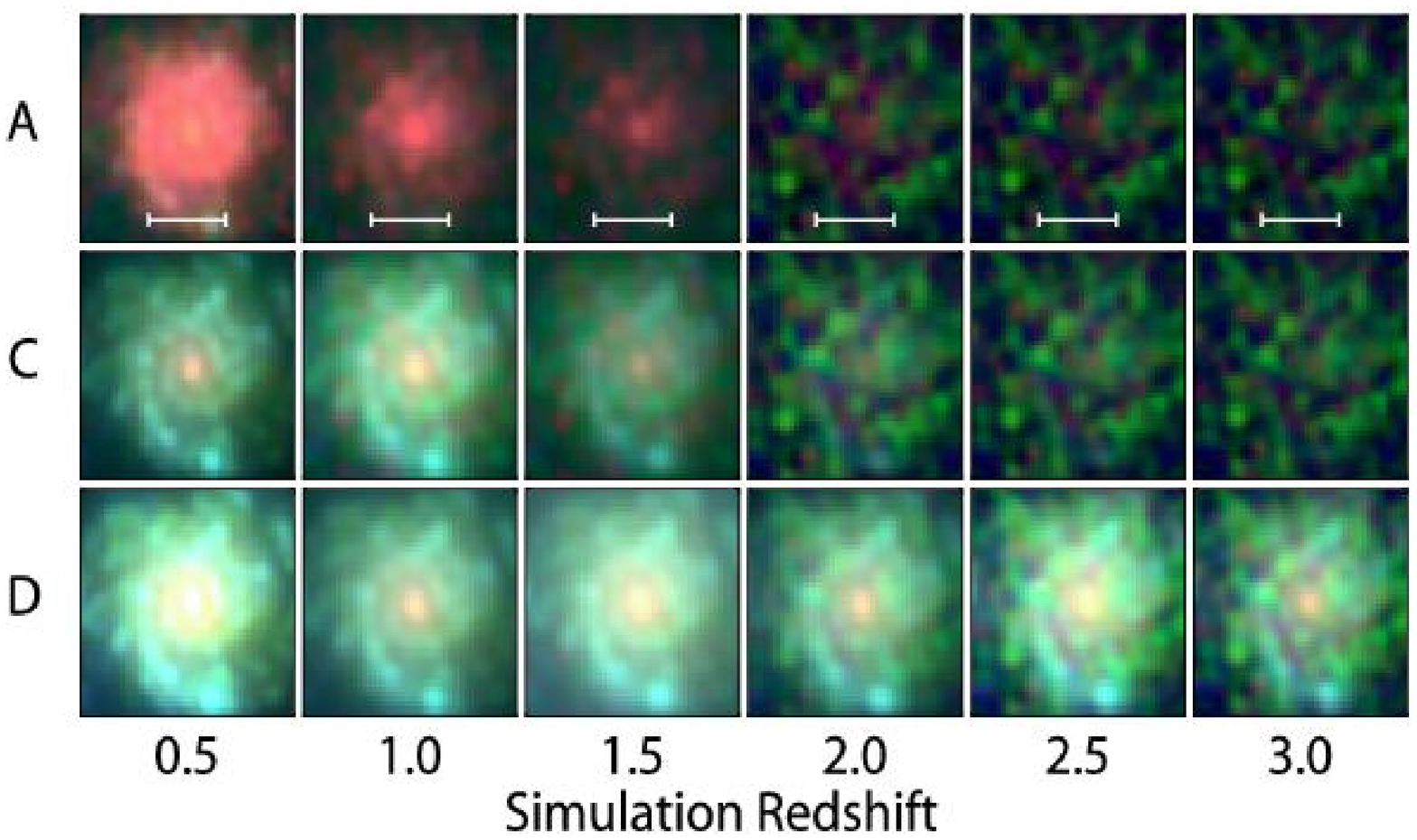}
\caption{Same as figure~\ref{fig:zsimngc3031}, but for NGC~628.  Note
that for model (D), the luminosity of NGC~628 has been increased by
2~mag in each of the FUV--, MUV--, and $B$--bands.  The horizontal bar
in each panel of the top row corresponds to a distance of 5~kpc in the
rest frame of the galaxy at each redshift.}\label{fig:zsimngc628}
\end{figure}

\begin{figure}
\plotone{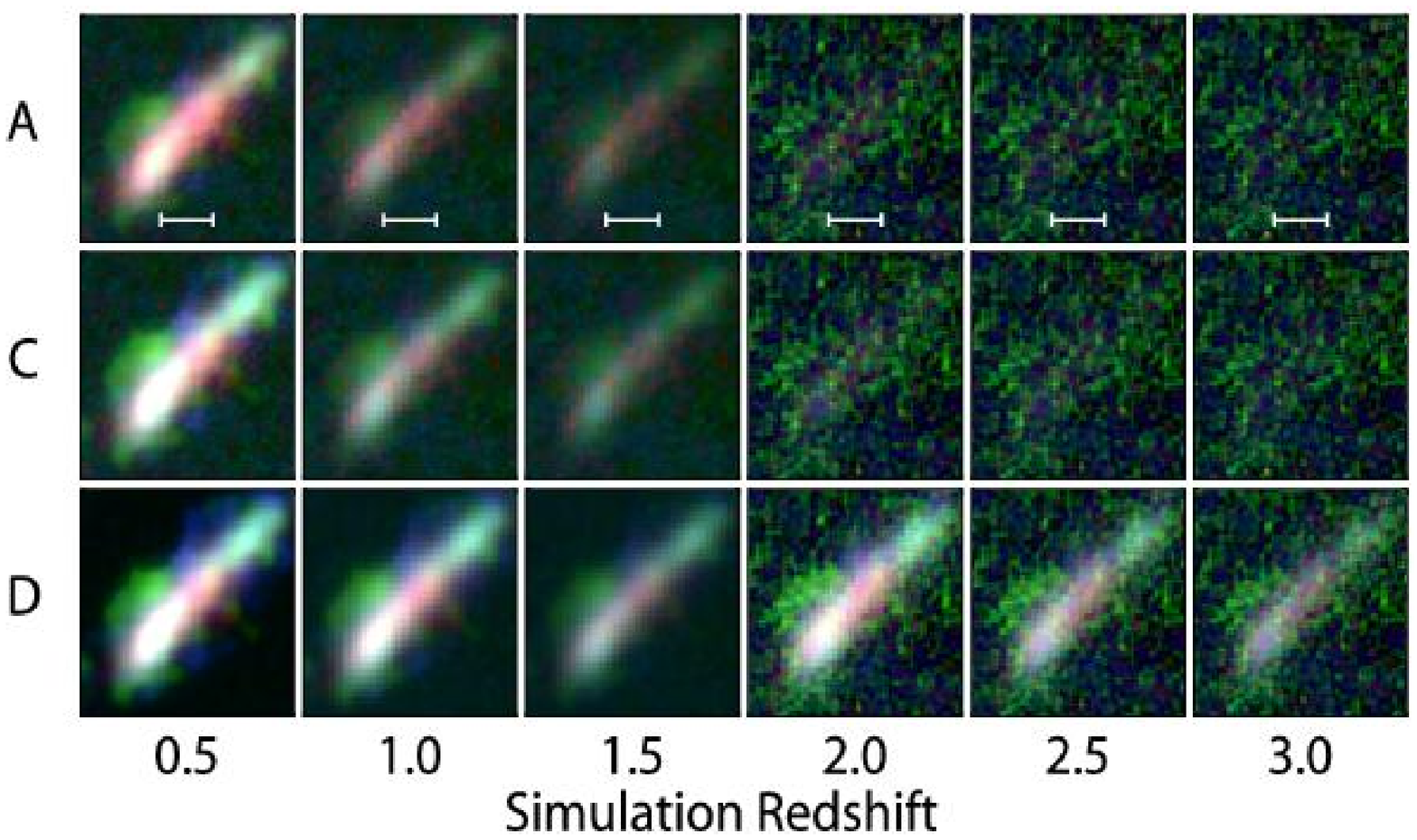}
\epsscale{1.0}
\caption{Same as figure~\ref{fig:zsimngc3031}, but for UGC~06697.
Note that for model (D), the luminosity of UGC~06697 has been
increased by 2~mag in each of the FUV--, MUV--, and $B$--bands.  The
horizontal bar in each panel of the top row corresponds to a distance
of 10~kpc in the rest frame of the galaxy at each
redshift.}\label{fig:zsimugc6697} 
\end{figure}

Figures~\ref{fig:zsimngc3031}--\ref{fig:zsimugc6697} illustrate the
appearance of NGC~3031, NGC~628, and UGC~06697, respectively, under
the evolution models of no evolution (A), pure color evolution with no
change in luminosity (C), and color evolution and brightening of
luminosity (D), as marked in each figure (see discussion above).  The
internal color dispersion of NGC~628 is reduced after convolution of
the NICMOS PSF and addition of the \hdf\ backgrounds.  However, the
resulting values are significant [$\xi(\mathrm{MUV},B) \simeq 0.1$;
$\xi(\mathrm{FUV},\mathrm{MUV}) \lsim 0.05$], and these are consistent
with values derived for other potentially large spiral galaxies at
$z\sim 1$ in the \hdf\ data (see Paper~II). In the case of UGC~06697,
which is already relatively blue in its UV--to--optical colors,
surface brightness dimming makes the galaxy essentially undetectable
for $z\gsim 2$.  However, by brightening the galaxy in luminosity (by
two magnitudes) its morphology is easily discernible to $z = 3$.  The
patchy star--formation, which presumably is responsible for the
observed internal color dispersion, also produces a measurable signal
out to this redshift.  Strong internal color dispersion is measured in
NGC~3031 under luminosity evolution combined with color evolution,
which appears to derive entirely from largely segregated stellar
populations that produce the red bulge and blue spiral arm colors.  If
such systems exist at $z\sim 2-3$ with the luminosities and colors of
observed galaxies at these redshifts, then they would be detectable
with sufficient S/N for quantitative morphological analysis.

One interesting conclusion from these simulations is that even under
the assumption of no evolution in the (rest--frame) $B$--band
luminosities the \uit\ galaxies would be detectable in deep NICMOS
images (like those for the \hdf). The limiting $\nich$ magnitude
($10\sigma$) for the \hdf\ data is $H_{160,\mathrm{lim}} \simeq 26.5$
\citep{dic02}.  Taking the distance modulus for a galaxy to
$z\gsim 2$, this limiting magnitude corresponds roughly to an
absolute, $B$--band magnitude, $M_{B_0,\mathrm{lim}} \sim -19$, in the
rest--frame (although this neglects surface--brightness dimming
effects on aperture measurements).  The majority of the galaxies in
the \uit\ sample have luminosities in excess of this limit, and would
be detected if present even out to these high redshifts.  In contrast,
our simulations agree with previous statements that the UV
luminosities are generally too faint to be visible for redshifts much
greater than unity \citep[see also, \eg,][]{gia95,kuc00,win02}.

\section{Summary}

We have discussed an indicator of the internal color dispersion within
galaxies, which has quantified the morphological $K$--correction,
namely, the dependence of galaxy morphology with observed wavelength.
We have measured the UV--optical internal color dispersion of a sample
of local galaxies using archival \uit\ data at 1500~\AA\ and 2500~\AA,
and ground--based $B$--band images.  Our findings show that the
internal color dispersion $\xi$ between the UV and optical passbands
is sensitive to the distribution of the stars that emit at these
wavelengths, and specifically that it gauges the spatial heterogeneity
of these stellar populations.

The mid--type spiral galaxies in the \uit\ sample generally exhibit
the highest dispersion in their ultraviolet--to--optical internal
colors, \eg, NGC 3031, NGC~4321, NGC~628.  The high dispersion
apparently results from differences in the stellar content that
constitute the bulge, disk, and spiral--arm components.  In the case
of NGC~3031 (one of the nearest examples of a ``grand--design''
spiral), the bulge--spiral-arm dichotomy clearly dominates the derived
internal color dispersion between both the \uit/MUV and $B$--bands,
and between the \uit/FUV and \uit/MUV images.  This illustrates the
effects that segregated stellar populations of varying ages (older
bulge populations versus the star--forming spiral arms) have on the
dispersion of the internal UV--to--optical colors.

Irregulars and later--type spirals also show significant internal
color dispersion, \eg, NGC~2146, NGC~3034, UGC~06697, although with
values that are generally lower than that of the mid--type spirals.
The smaller $\xi$ values presumably arise from the presence of young
stellar populations that dominate the flux emission at UV--to--optical
wavelengths, which are modulated by (varying) dust absorption.

Ellipticals, lenticulars, and early--type spirals generally have low
or negligible internal color dispersion, as evidenced by NGC~1316, NGC
1399, and NGC~4486.  The general result, given that early--type galaxies have
small internal color dispersion coupled with the relative
absence of ongoing star formation, is that older well--mixed
stellar populations dominate the flux at UV--to--optical wavelengths.

There is little dependence between the UV--optical total galaxy colors
and $\xi(\mathrm{FUV,MUV})$ for the \uit\ sample.
However, the $\xi(\mathrm{MUV},B)$ values exhibit some
general trends with the integrated $m_{2500} - B$ colors.   Galaxies
with the bluest and reddest colors, \ie, $m_{2500} - B \lsim 2$ and
$m_{2500}-B \gsim 3$, generally have lower $\xi(\mathrm{MUV},B)$
values, with smaller scatter.  For galaxies with the bluest colors, a
single (young) stellar population likely dominates the light at
UV--optical colors, which amounts to little internal color dispersion.
The (small) internal color dispersion values in such systems is likely
caused by dust absorption variations.  Likewise, the stellar
populations that dominate these colors in galaxies with $m_{2500} - B
\gsim 3$ must be relatively co-spatial.  Galaxies with intermediate
MUV--optical colors, $2 \lsim (m_{2500} - B) \lsim 3$, show a
very large scatter in their $\xi(\mathrm{MUV},B)$ values.

The scatter in $\xi(\mathrm{MUV},B)$ for intermediate MUV--optical
colors hints at the physical conditions that produce high
$\xi(\mathrm{MUV},B)$, and likely implies that these colors are in a
sense \textit{necessary}. We have used simple simulations to
understand how a mix of young and old stellar populations affects the
MUV--optical internal color dispersion.  We find that high
$\xi(\mathrm{MUV},B)$ results only in the case that a small amount of
the total galaxy stellar mass is in young stellar populations, \ie,
$\sim 0.01 - 10$\% by mass.  In this range of stellar--mass fractions,
the young stellar populations dominate the MUV flux emission, while
the older stellar populations contribute substantially at optical
wavelengths.  At smaller fractions, the older stars begin to dominate
the MUV emission, and at higher stellar--mass fractions, the young
stellar populations dominate the flux from UV--to--optical wavelengths.

We find an interesting relationship between the $\xi(\mathrm{MUV},B)$
values and the optical--to--FIR flux ratios,
$f_\nu(B)/f_\nu(60\micron)$, in the \uit\ sample.  Galaxies with
higher $f_\nu(B)/f_\nu(60\micron)$ show an increased upper envelope
and scatter in $\xi(\mathrm{MUV},B)$ relative to galaxies with lower
optical--to--FIR flux ratios.  The reason for this is not completely
clear.  However, because galaxies with lower optical--to--FIR flux
ratios show evidence for heavy dust absorption and span a smaller
range of $\xi(\mathrm{MUV},B)$ values, this seems to imply that our
internal color dispersion statistic is more sensitive to variations in
the galaxies' stellar populations rather than to dust absorption.  To
investigate this further will require future galaxy surveys that cover
UV--to--FIR wavelengths for large galaxy samples, which should be
possible with planned surveys with \textit{SIRTF}.

We have simulated the appearance of local galaxies from the \uit\
sample at cosmological distances ($z = 0.5-3.0$) to investigate the
application of internal color dispersion statistic to high--redshift
galaxies in deep, \hst\ images. Many of the \uit--sample galaxies have
luminosities that are sufficiently bright in the rest--frame optical
to be detected within the limits of the currently deepest
near--infrared surveys even with no evolution.  Thus, large,
present--day galaxies (if indeed they were present) would be detected
in these surveys.  However, we find (in agreement with previous
studies) that the \uit\ galaxies generally have UV luminosities too
faint for detection at high redshift ($z\gsim 2$) even out to the
limits of the deepest \hst\ images, and far below the S/N requirements
for reliable quantitative morphological analysis.  However, under
basic assumptions that the luminosities and UV--to--optical colors of
the local galaxies are comparable to the values of high--redshift
objects, we have shown that the galaxies' derived internal color
dispersion remains measurable out to $z\sim 3$.

In a forthcoming paper (Paper~II), we apply our internal color
dispersion statistic to samples of high--redshift galaxies at $z\sim
1$ and $z\sim 2.7$ using \hst\ data available for the \hdf.  As we
have demonstrated here, this statistic is highly useful for
identifying galaxies with both segregated stellar populations that
dominate to different wavelengths of the galaxies' SEDs, and variable,
``patchy'' dust opacity.  The \uit\ sample studied here will thus
provide a useful benchmark for these high redshift comparisons for
interpretation of the assembly and evolution of the stellar content in
these galaxies over time.

\acknowledgements

We wish to thank our colleagues at the Steward Observatory and the
Space Telescope Science Institute for stimulating conversations. In
particular, we would like to acknowledge helpful comments and
suggestions from Eric Bell, Romeel Dav\'e, Elizabeth Barton Gillespie,
Katherine Hu, Jennifer Lotz, Matthias Steinmetz, and Andrew Zirm.  We
also wish to thank the anonymous referee whose comments enhanced the
presentation of this paper.  This research has made use of the
NASA/IPAC Extragalactic Database (NED) which is operated by the Jet
Propulsion Laboratory, California Institute of Technology, under
contract with the National Aeronautics and Space Administration.


\appendix
\section{Simulating the Appearance of Galaxies at Cosmological 
Distances\label{section:ztrans}}

To simulate galaxy properties shifted to more distant reference frames
in an expanding Universe, we have generally followed the prescription
outlined by \citet{gia95}.  This method implicitly assumes zero
evolution between the galaxies' appearances.  The original images are
first converted to physical flux density units (flux per unit
frequency, $f_\nu$).  The images are then convolved with a kernel such
that final rebinned image matches the required PSF.  Shifting the
galaxy images to cosmological distances imposes two effects.  First,
the apparent size of the galaxy changes via the angular diameter
distance (\ie, the angular size of a fixed rod at redshift, $z$).
Thus, when one observes a galaxy with angular size, $\theta_{1}$, and
distance, $D_{1}$, shifted to a new cosmological distance, its
apparent angular size subtends an angle, $\theta_{2} =\theta_{1}\,
D_{A_1} / D_{A_2}$, where  $D_A$ is the angular diameter distance (and
depends on the underlying cosmology).  The relation between the
observed size of a galaxy at a distance, $D_1$, imaged with a detector
with a physical pixel scale, $p_{1}$, and the size when subsequently
imaged at a distance, $D_2$, with an intrinsic pixel scale, $p_{2}$,
can then be expressed as,
\begin{equation}\label{eqn:rebin}
\frac{\theta_{1}}{p_{1}} = b\, \frac{\theta_{2}}{p_{2}}.
\end{equation}
The constant, $b$, is the factor by which the image of the galaxy must
be ``rebinned'' to simulate its observed appearance at higher redshift
in the given detector.  Rewriting equation~\ref{eqn:rebin} then
yields, $b = (D_{A_2}/D_{A_1}) (p_{2}/p_{1})$.  This  rebinning factor
is applied to each  galaxy image, averaging over pixel intensity
values in order to preserve the object surface brightness, which is
independent of cosmology.  To do the rebinning, we have used the IDL
function FREBIN, which has the property that each input pixel
is equally represented in the output matrix (\ie, flux is preserved).
This algorithm sums over the flux contributions from the fraction of
each input pixel within the boundary of the output pixel.  Thus, the
total image flux or surface brightness can be conserved (the latter
where each output pixel is normalized by the pixel area).

The second effect imposed by shifting reference frames derives from
the relativistic cosmological expansion.  For example, the surface
brightness is defined as the energy from an object emitted
per unit time per unit area per area of the
incident detector (\ie, the flux per unit area).  The energy,
time, and area all transform with the cosmological redshift.  Counting
``powers'' of $(1+z)$, the energy of an incident photon is reduced by
one power, and the length of unit time and each of the two dimensions
of the unit area are each stretched by one power.  Thus, the net
affect of the cosmological expansion is to attenuate the observed
bolometric surface brightness by $\propto (1+z)^{-4}$.   To
account for these effects we have utilized the property that the
intrinsic galaxy luminosity (the flux integrated over the surface
area) is preserved.  This leads to the equation,
\begin{equation}\label{eqn:lumcons}
4\pi D_{L_1}^2\, N_{1}\, f_\nu(z_1)\,(1+z_1)^{-1} = 4\pi D_{L_2}^2\,
	N_{2}\, f_\nu(z_2)\, (1+z_2)^{-1},
\end{equation}
where, $f_\nu(z)$, is the galaxy's flux density averaged over the
number of pixels, $N$ (with corresponding area, $p^2$, per pixel),
and $D_L$ is the luminosity distance.  Note that for local galaxies,
$z_1$ is nil.  The observed and redshifted bandpasses here have
generally been selected such that $\lambda_{z_2} = \lambda_{z_1}\,
(1+z_2)/(1+z_1)$ [due to the fact that the cosmological  expansion
degrades the energy of an observed photon by the factor of
$(1+z)^{-1}$].  Thus, this makes the approximation that the two
passbands have widths that sample co-moving wavelength ranges in the
observed frame, \ie, $\Delta \lambda_{z_2} \approx \Delta
\lambda_{z_1}\, (1+z_2)/(1+z_1) $.  Therefore,
equation~\ref{eqn:lumcons} becomes,
\begin{equation}
f_\nu(z_2) = f_\nu(z_1)\, \left( \frac{p_{2}}{p_{1}} \right)^2\,
[(1+z_1)/(1+z_2)]^{3},
\end{equation}
where this has utilized the facts that the physical pixel scale
corresponds  to a physical size on the sky by $p_{2} = \theta_{2}
/ N_{2} =(D_{A_1}/D_{A_2})\,/ N_{2}$, and the angular diameter
distance is related to the luminosity distance by $D_A = D_L\,
(1+z)^{-2}$.



\begin{thebibliography}{}

\bibitem[Abraham \etal(1999)]{abr99}
Abraham, R.~G., Ellis, R.~S., Fabian, A.~C., Tanvir, N.~R., \&
Glazebrook, K.\ 1999, \mnras, 303, 641 

\bibitem[Abraham \etal(1996)]{abr96}
Abraham, R.~G., Tanvir, N.~R., Sanitago, B.~X., Ellis, R.~S.,
Glazebrook, K., \& van den Bergh, S.\ 1996, \mnras, 279, 47 

\bibitem[Bahcall, Guhathakurta, \& Schneider(1990)]{bah90}
Bahcall, J.~N., Guhathakurta, P., \& Schneider, D.~P.\ 1990, Science, 248, 178

\bibitem[Bell \& de Jong(2001)]{bel01b} Bell, E.~F.~\& de 
Jong, R.~S.\ 2001, \apj, 550, 212 

\bibitem[Bell \& Kennicutt(2001)]{bel01}
Bell, E.~F., \& Kennicutt, R.~C., Jr.\ 2001, \apj, 548, 681

\bibitem[Bendinelli \etal(1990)]{ben90} Bendinelli, O., Zavatti, F., 
Parmeggiani, G., \& Djorgovski, S.\ 1990, \aj, 99, 774 

\bibitem[Bernstein, Freedman, \& Madore(2002)]{bur02} 
Bernstein, R.~A., Freedman, W.~L., \& Madore, B.~F.\ 2002, \apj, 571, 107 

\bibitem[Bertin \& Arnouts(1996)]{ber96}
Bertin, E., \& Arnouts, S.\ 1996, \aaps, 117, 393

\bibitem[Blanton et al.(2001)]{bla01} Blanton, M.~R.~et al.\ 
2001, \aj, 121, 2358 

\bibitem[Bohlin \etal(1991)]{boh91} Bohlin, R.~C.~et al.\ 
1991, \apj, 368, 12 

\bibitem[Brinchmann \etal(1998)]{bri98}
Brinchmann, J., \etal\ 1998, \apj, 499, 112

\bibitem[Brown \etal(2000)]{bro00}
Brown, T.~M., Bower, C.~V., Kimble, R.~A.~, Sweigart, A.~V., \&
Ferguson, H.~C.\ 2000, \apj, 532, 308 

\bibitem[Bruzual \& Charlot(1993)]{bru93}
Bruzual, G.~A., \& Charlot, S.\ 1993, \apj, 405, 538

\bibitem[Calzetti(2001)]{cal01} 
Calzetti, D.\ 2001, \pasp, 113, 1449 

\bibitem[Calzetti \etal(2000)]{cal00} Calzetti, D., Armus, L., Bohlin,
R.~C., Kinney, A.~L., Koornneef, J., \& Storchi--Bergmann, R.\ 2000,
\apj, 533, 682

\bibitem[Calzetti, Kinney, \& Storchi--Bergmann(1994)]{cal94} 
Calzetti, D., Kinney, A.~L., \& Storchi--Bergmann, T.\ 1994, \apj, 429, 582

\bibitem[Cardelli, Clayton, \& Mathis(1989)]{car89} Cardelli, J.~A.,
Clayton, G.~C., \& Mathis, J.~S.\ 1989, \apj, 345, 245  

\bibitem[Cheng \etal(1996)]{che96}
Cheng, K.--P., Hintzen, P., Smith, E.~P., Angione, R., Talbert, F.,
Collins, N., \& Stecher, T.\ 1996, Ground--Based CCD Imaging in Support
of the \uitastro{1}/\uit\ Space Shuttle Mission (3 vols.; NSSDC CD--ROM) 

\bibitem[Conselice \etal(2000)]{con00}
Conselice, C.~J., Bershady, M.~A., \& Jangren, A.\ 2000, \apj, 529, 886


\bibitem[Conselice, Gallagher, \& Wyse(2001)]{con01} 
Conselice, C.~J., Gallagher, J.~S., \& Wyse, R.~F.~G.\ 2001, \aj, 122, 2281 

\bibitem[Conselice(2003)]{con03}
Conselice, C.~J., 2003, \apjs, 147, 1

\bibitem[de Jong(1996)]{dej96} de Jong, R.~S.\ 1996, \aap, 
313, 377 

\bibitem[de~Vaucouleurs \etal(1991)]{dev91}
de Vaucouleurs, G., de Vaucouleurs, A., Corwin Jr., H.G., Buta, R.J.,
Fouque, P., \& Paturel, G.\ 1991 The Third Reference Catalogue of
Bright Galaxies, (New York: Springer--Verlag) [RC3] 

\bibitem[Dickinson(1999)]{dic99} 
Dickinson, M.\ 1999, in \textit{After the Dark Ages: When Galaxies
  were Young (the Universe at $2 < z < 5$)}, eds.\ S.~Holt, \&
E.~Smith (New York, American Institute of Physics), 122 


\bibitem[Dickinson \etal(2003a)]{dic02b}
Dickinson, M., Papovich, C., Ferguson, H.~C., \& Budav\'ari, T.\ 2003a,
\apj, 587, 25

\bibitem[Dickinson \etal(2003b)]{dic02} Dickinson, M., 2003b, in preparation

\bibitem[Dorman, 0'Connel, \& Rood(1995)]{dor95}
Dorman, B., O'Connel, R.~W., \& Rood, R.~T.\ 1995, \apj, 442, 105

\bibitem[Elmegreen \etal(1996)]{elm96}
Elmegreen, B.~G., Elmgreen, D.~M., Salzer, J.~J., \& Mann, H.\ 1996,
\apj, 467, 579

\bibitem[Elmegreen \etal(1994)]{elm94}
Elmegreen, D.~M., Elmegreen, B.~G., Lang, C., \& Stephens, C.\ 1994,
\apj, 425, 57

\bibitem[Eskridge, \etal(2003)]{esk03}
Eskridge, P.~B.\ \etal\ 2003, \apj, 586, 923

\bibitem[Ferrarese \etal(1996)]{fer96}
Ferrarese, L.~\etal\ 1996, \apj, 464, 568.

\bibitem[Freedman \etal(1994)]{fre94}
Freedman, W.~L.~\etal\ 1994, ApJ, 417, 680

\bibitem[Freedman et al.(2001)]{fre01} Freedman, W.~L.~et al.\ 2001,
  \apj, 553, 47 

\bibitem[Fruchter \& Hook(2002)]{fru02} Fruchter, A.~S.~\& 
Hook, R.~N.\ 2002, \pasp, 114, 144 

\bibitem[Gadotti \& dos Anjos(2001)]{gad01}
Gadotti, D.~A., \& dos Anjos, S.\ 2001, \aj, 122, 1298
 
\bibitem[Gavazzi \etal(1998)]{gav98} Gavazzi, G., Catinella, 
B., Carrasco, L., Boselli, A., \& Contursi, A.\ 1998, \aj, 115, 1745 

\bibitem[Giavalisco \etal(1995)]{gia95}
Giavalisco, M., Livio, M., Bohlin, R.~C., Macchetto, F.~D., \&
Stecher, T.~P.\ 1995, \aj, 112, 369 

\bibitem[Giavalisco, Steidel, \& Macchetto(1996)]{gia96}
Giavalisco, M., Steidel, C.~C., \& Macchetto, F.~D.\ 1996, \apj, 470, 189 

\bibitem[Hutchings et al.(1990)]{hut90}
Hutchings, J.~B., Lo,  E., Neff, S.~G., Stanford, S.~A., \& Unger,
S.~W.\ 1990, \aj, 100, 60 

\bibitem[Kohle \etal(1996)]{koh96}
Kohle, S., Kissler--Patig, M., Hilker, M., Richtler, T., Infante, L.,
\& Quintana, H.\ 1996, \aap, 309, 39 
 
\bibitem[Kron(1980)]{kro80}
Kron, R.~G.\ 1980, \apjs, 43, 305

\bibitem[Kuchinski \etal(2001)]{kuc01}
Kuchinski, L.~E., Mardore, B.~F., Freedman, W.~L., \& Trewhella,
M.\ 2001, \aj, 122, 729

\bibitem[Kuchinski \etal(2000)]{kuc00}
Kuchinski, L.~E., \etal\ 2000, \apjs, 131, 441

\bibitem[Lauer(1999)]{lau99} 
Lauer, T.~R.\ 1999, \pasp, 111, 1434 

\bibitem[Lilly \etal(1998)]{lil98}
Lilly, S.~J., \etal\ 1998, \apj, 500, 75

\bibitem[Lowenthal \etal(1997)]{low97} Lowenthal, J.~D., \etal\ 1997,
\apj, 481, 673.

\bibitem[Marcum \etal(2001)]{mar01}
Marcum, P.~M., \etal\ 2001, \apjs, 132, 129

\bibitem[Marleau \& Simard(1998)]{mar98}
Marleau, F.~R., \& Simard, L.\ 1998, \apj, 507, 585

\bibitem[McMillan, Ciardullo, \& Jacoby(1993)]{mcm93} 
McMillan, R., Ciardullo, R., \& Jacoby, G.~H.\ 1993, \apj, 416, 62 

\bibitem[McNamara \& O'Connell(1989)]{mcn89} McNamara, 
B.~R.~\& O'Connell, R.~W.\ 1989, \aj, 98, 2018 

\bibitem[Menanteau, Abraham, \& Ellis(2001)]{men01}
Menanteau, F., Abraham, R.~G., \& Ellis, R.~S.\ 2001, \mnras, 322, 1

\bibitem[Moshir et al.(1990)]{mos90} Moshir, M.~et al.\ 1990, IRAS
  Faint Source Catalogue, version 2.0 (1990) 

\bibitem[Neff et al.(1994)]{nef94} Neff, S.~G., Fanelli, 
M.~N., Roberts, L.~J., O'Connell, R.~W., Bohlin, R., Roberts, M.~S., Smith, 
A.~M., \& Stecher, T.~P.\ 1994, \apj, 430, 545 

\bibitem[Ohl et al.(1998)]{ohl98} Ohl, R.~G.~et al.\ 1998, 
\apjl, 505, L11 

\bibitem[Oke, Gunn, \& Hoessel(1996)]{oke96}
Oke, J., Gunn, J., \& Hoessel, J.\ 1996, \aj, 111, 29


\bibitem[Papovich, Dickinson, \& Ferguson(2001)]{pap01}
Papovich, C., Dickinson, M., \& Ferguson, H.~C.\ 2001, \apj, 559, 620

\bibitem[Papovich \etal(2003)]{pap02b}
Papovich, C., Dickinson, M., Giavalisco, M., Conselice, C.~C., \&
Ferguson, H.~C.\ 2003, in preparation (Paper II)
 
\bibitem[Rice et al.(1988)]{ric88} Rice, W., Lonsdale,
C.~J., Soifer, B.~T., Neugebauer, G., Koplan, E.~L., Lloyd, L.~A., de
Jong, T., \& Habing, H.~J.\ 1988, \apjs, 68, 91 

\bibitem[Roberts \& Haynes(1994)]{rob94}
Roberts, M.~S., \& Haynes, M.~P.\ 1994, \araa, 32 115

\bibitem[Roberts \etal(2001)]{rob01}
Roberts, T.~P., Schurch, N.~J., \& Warwick, R.~S. 2001, \mnras, 324,
737 

\bibitem[Romaniello \etal(2002)]{rom02}
Romaniello, M., Panagia, N., Scuderi,  S., \& Kirshner, R.~P.\ 2002,
\aj, 123, 915 

\bibitem[Schade \etal(1995)]{sch95}
Schade, D., Lilly, S.~J., Crampton, D., Hammer, F., Le Fevre, O.,
\& Tresse, L.\ 1995, \apj, 451, L1 

\bibitem[Schlegel, Finkbeiner, \& Davis(1998)]{sch98}
Schlegel, D.~J, Finkbeiner, D.~P., \& Davis, M.\ 1998, ApJ, 500, 525

\bibitem[Schweizer(1980)]{sch80} 
Schweizer, F.\ 1980, \apj, 237, 303 

\bibitem[Sharina \etal(1996)]{sha96}
Sharina, M.~E., Karachentsev, I.~D., \& Tikhonov, N.~A.\ 1996, \aaps,
119, 499 

\bibitem[Shaya et al.(1996)]{shaya96} 
Shaya, E.~J.~et al.\ 1996, \aj, 111, 2212 

\bibitem[Simard \etal(1999)]{sim99}
Simard, L., \etal\ 1999, \apj, 519, 563

\bibitem[Stecher \etal(1997)]{ste97}
Stecher, T.~P., \etal\ 1997, \pasp, 109, 584

\bibitem[Trager \etal(2000)]{tra00} Trager, S.~C., Faber, S.~M.,
  Worthey, G., \& Gonz{\' a}lez, J.~J.\ 2000, \aj, 120, 165

\bibitem[van den Bergh(2002)]{vdb02} 
van den Bergh, S.\ 2002, \pasp, 114, 79

\bibitem[van den Bergh, \etal(1996)]{van96}
van den Bergh, S., Abraham, R.~G., Ellis, R., S., Tanvir, N.~R., \&
Glazebrook, K.~G.\ 1996, \aj, 112, 359 

\bibitem[van den Bergh, Cohen, \& Crabbe(2001)]{vdb01} van 
den Bergh, S., Cohen, J.~G., \& Crabbe, C.\ 2001, \aj, 122, 611

\bibitem[Williams \etal(1996)]{wil96}
Williams, R.~E., \etal\ 1996, \aj, 112, 1335

\bibitem[Windhorst \etal(2002)]{win02}
Windhorst, R.~A., \etal\ 2002, \apjs, 143, 113

\end{thebibliography}
\end{document}